\documentclass[pra,aps,twocolumn,showpacs,superscriptaddress,floatfix,amsmath,amssymb,nofootinbib]{revtex4}
\newcommand{\beq}{\begin{equation}}
\newcommand{\eeq}{\end{equation}}
\newcommand{\beqa}{\begin{eqnarray}}
\newcommand{\eeqa}{\end{eqnarray}}

\def\ra{\rangle}
\def\la{\langle}

\usepackage{graphicx}
\usepackage{bm}
\usepackage{pdfpages}

\usepackage{color}

\begin{document}
\title{Adiabaticity condition for non-Hermitian Hamiltonians}

\author{S. Ib\'a\~{n}ez}
\affiliation{Departamento de Qu\'{\i}mica F\'{\i}sica, Universidad del Pa\'{\i}s Vasco - Euskal Herriko Unibertsitatea,
Apdo. 644, Bilbao, Spain}

\author{J. G. Muga}
\affiliation{Departamento de Qu\'{\i}mica F\'{\i}sica, Universidad del Pa\'{\i}s Vasco - Euskal Herriko Unibertsitatea,
Apdo. 644, Bilbao, Spain}
\affiliation{Department of Physics, Shanghai University, 200444 Shanghai, China}
\begin{abstract}
We generalize the concept of population for non-Hermitian systems in different ways and identify the one 
best suited to characterize adiabaticity. {An approximate adiabaticity criterion consistent with this choice is also 
worked out.} Examples are provided for different processes involving two-level atoms with decay.   
\end{abstract}
%32.80.Qk.Coherent control of atomic interactions with photons
%42.50.-p Quantum Optics
\maketitle
\section{Introduction}
The adiabatic approximation and adiabatic following are key concepts to study and manipulate quantum systems.
%as well as electromagnetic fields in waveguides.  
%,  i.e., atoms, ions or molecules, being fundamental in quantum optics and ultracold atom physics.
%The different processes applied to these systems, such as cooling, transport, condensing, splitting, etc., must be carried out slow enough so that the internal states of the systems are not excited, avoiding the heating.
For a  time-dependent Hamiltonian, the instantaneous eigenvalues  and their corresponding eigenstates
change with time along time-dependent ``trajectories''.    
%These virtual trajectories  do not represent in general actual dynamical paths followed by the system since the Hamiltonian induces transitions between the instantaneous eigenstates. 
%However, 
For very slow changes of the control parameters the system will follow closely 
an  eigenstate trajectory  
up to a phase factor if it is initially in one of the eigenstates. This is the essence of ``adiabaticity'' \cite{BF},
and the adiabatic approximation provides the 
form of the phase\footnote{A broader concept of ``adiabaticity''  applies also to time independent 
Hamiltonians with several degrees of freedom,
for example in the Born-Oppenheimer approximation.}.  
Adiabaticity is useful for several reasons:  the phase factors accompanying adiabatic changes  \cite{berry1984} imply many consequences in 
atomic, molecular, optical, and condensed matter physics;  as well,  setting initial and final Hamiltonians 
as boundary conditions, 
%(which define the preparation initially and the target state(s) at the final time), 
the details of the parameter paths connecting them are unimportant for the final populations 
as long as the process is adiabatic. This feature explains the robustness of  
adiabatic methods to prepare  states; the robustness of adiabatic devices such as the atom diode \cite{ad1,ad2};  
or some applications of adiabaticity  in quantum information processing
%(In adiabatic computing the adiabatic evolution is used    
%to prepare an unknown final ground state 
%that encodes the solution of a computational problem starting from a ground initial state which is easy to prepare 
\cite{qc}.
%). 
%Adiabaticity is also instrumental in designing  ``quiet''  transitions that minimize the transient excitation energy.  
%As well, for a given time-dependent Hamiltonian and initial state the adiabatic approximation provides the minimal 
%instantaneous energies.  
Thus, knowing the conditions that determine the adiabaticity of a given process is generically worthwhile.  
For Hermitian Hamiltonians the conditions imply the conservation of the populations for the 
(time-dependent) instantaneous orthonormal eigenstates \cite{Schiff}. 
For non-Hermitian (NH) systems, however, the {``population''} concept is problematic because of the arbitrariness
in the normalization of right and left eigenvectors \cite{jolicard,BU,erratum},
and because of their non-orthogonality and the ensuing non-diagonal contributions to the total state norm. 
In addition, the usual approximations and criteria are not necessarily valid, 
%and NH degeneracies may appear,    
so arguments and results which are applicable for Hermitian systems have to be reconsidered and modified \cite{nenciu,ChangPu,guerin,jolicard,FM,sar_sof,erratum,Jolicard2012,GM}.
%\red{(**************** poner algo sobre least dissipative state??**********)}
In this paper we provide first different generalizations of the population concept for NH systems and examine their properties. 
We identify among them the one which is  best suited to {define the adiabaticity condition}. 
%, even if it 
%does not satisfy  the conditions 
%associated with an ordinary  population.  
% Thus, these adiabatic invariants can be understood as the conditions which determines the adiabaticity. This makes one wonder about the concept of adiabaticity for non-Hermitian systems. On the other hand, 
We then provide an approximate expression for the adiabaticity condition that {improves on previous proposals.}
Its limitations are also pointed out and examples are presented.    
%In this paper we shall write down adiabaticity conditions that are completely independent of the 
%eigenstate normalization. One is based on wavevector components and is independent of the concept of population. The other one generalizes the concept of population.  

Let us briefly 
review the relations that characterize 
a non-Hermitian system described by a time-dependent Hamiltonian $H(t)$ with $N$
non-degenerate right eigenstates $\{|n(t)\ra\}$, $n=1,2...,N$,
\cite{Muga}
\beq
\label{eigenvalues_eq}
H(t) |n(t)\ra = E_n(t) |n(t)\ra,
\eeq
and biorthogonal partners $\{|\widehat{n}(t)\ra\}$ which are left eigenstates. Equivalently, 
\beq
\label{eigenvalues_eq1}
H^{\dag}(t)  |\widehat{n}(t)\ra = E_n^{*}(t)  |\widehat{n}(t)\ra, 
\eeq
where the star means ``complex conjugate'' and the dagger denotes the adjoint
operator.   
They are normalized to satisfy the biorthogonality relation
\beq
\label{delta}
\la\widehat{n}(t)|m(t)\ra = \delta_{nm}
\eeq
and the closure relations
\beq
\label{closure}
\sum_n |\widehat{n}(t)\ra \la n(t)| = \sum_n |n(t)\ra\la \widehat{n}(t)|=1.
\eeq
%
%$\la\widehat{n}(t)|$ is the left eigenvector of $H(t)$,
%
%\beq
%\la\widehat{n}(t)| H(t)= \la\widehat{n}(t)| E_n(t),
%\eeq
%
%and $\la n(t)|$ the left eigenvector of $H^{\dag}(t)$,
%
%\beq
%\la n(t)| H^{\dag}(t)= \la n(t)|  E_n^{*}(t).
%\eeq
%
%We can thus write the Hamiltonian and its adjoint as
%
%\beqa
%\nonumber
%H(t)&=& \sum_n |n(t)\ra E_n(t) \la\widehat{n}(t)|,
%\\
%H^{\dag}(t) &=& \sum_n |\widehat{n}(t)\ra E_n^{*}(t) \la n(t)|.
%\eeqa
%
%The time-dependent Schr\"{o}dinger equations for
%a generic state $|\Psi(t)\ra$ and for its biorthogonal partner
%$|\widehat{\Psi}(t)\ra$
%satisfying $\la \widehat{\Psi}(t)|\Psi(t)\ra=1$ are
%
%
%and $\Psi(0)=\hat{\Psi(0)}$
%
%\beqa
%\label{Schrodinger_eq}
%i\hbar |\dot{\Psi}(t)\ra&=& H(t) |\Psi(t)\ra,
 %\\
%\label{adjoint_eq}
%i\hbar |\dot{\widehat{\Psi}}(t)\ra&=& H^{\dag}(t) |\widehat{\Psi}(t)\ra,
%\eeqa
%
%{where the dot represents derivative with respect to time.} 
The states
\beqa
\label{set_eigenstates1}
|\phi_n(t)\ra &=& f_n(t) |n(t)\ra,
\\ 
|\widehat{\phi}_n(t)\ra &=& \frac{1}{f_n^*(t)} |\widehat{n}(t)\ra,
\label{set_eigenstates2}
\eeqa
where $f_n(t) \in \mathbb{C}$ is an arbitrary function \cite{jolicard}, 
constitute also a complete, biorthogonal set of eigenstates of $H(t)$. Thus, 
the freedom to define the eigenvectors of NH Hamiltonians
goes  
beyond the imaginary phase factor ambiguity of 
Hermitian ones and their ordinary norm $\sqrt{\la n(t)|n(t)\ra}$ can be arbitrary. 
Some restrictions on the $f_n(t)$ apply if the basis is parallel transported, i.e., when $\la \widehat{n}(t)| \dot{n}(t) \ra=0$, where the dot means time derivative. 
From Eq. (\ref{set_eigenstates1}), taking into account Eq. (\ref{delta}), we find that
\beq
\label{cambio}
\la \widehat{\phi}_n(t)| \dot{\phi}_n(t) \ra = \dot{f}_n(t)/f_n(t) + \la \widehat{n}(t)| \dot{n}(t) \ra.
\eeq
Thus, if the reference basis $\{|n(t) \ra\}$ is parallel transported,
and the new basis should be parallel transported too, $f_n(t)$ must be constant. 
In other words, there is only one parallel transported basis
for each set of initial values $f_n(0)$, {where we fix $t=0$ as the initial time of the processes}. 
This will be useful later on.  

We may expand a state $|\Psi(t)\ra$ that satisfies the Schr\"odinger equation
\beq
\label{Schr_eq}
i\hbar |\dot{\Psi}(t)\ra = H(t) |\Psi(t)\ra,
\eeq
as 
\beqa
\label{psi1}
|\Psi(t)\ra = \sum_n c_n(t) |n(t)\ra.
\eeqa
%\\ 
%\label{psi2}
%&=& \sum_n \widetilde{c}_n(t) |\phi_n(t)\ra,
%\eeqa
%
From Eq. (\ref{closure}), $c_n(t)=\la \widehat{n}(t)|\Psi(t)\ra$,
% and $\widetilde{c}_n= c_n/f_n$, 
but the $|c_n|^2$  are not necessarily bounded by one, 
and their sum does not have to be one either. We may now explore the use of a convenient basis,
in particular regarding the definition of adiabaticity. 
  
A state with initial condition 
$|\Psi(0)\ra=|n(0)\ra$ behaves adiabatically if   
its dynamics is well approximated by  $e^{i\beta_n(t)}|n(t)\ra$. 
Substituting this  form as an ansatz into the Schr\"odinger equation (\ref{Schr_eq}) gives
\beq
\label{beta}
\beta_n(t)=-\frac{1}{\hbar} \int_0^t E_n(t') dt' +i \int_0^t \la \widehat{n}(t')|\dot{n}(t')\ra dt'. 
\eeq
For a general state, fully adiabatic dynamics (for all modes) would correspond to an evolution of the form  
\beq
\label{ansatz_betaad}
|\Psi(t)\ra= \sum_n g_n(0)|\psi_n(t)\ra, 
\eeq
where
\beqa
g_n(t)&:=&\la \widehat{\psi}_n(t)|\Psi(t)\ra,
\\ 
|\psi_n(t)\ra&:=&e^{i\beta_n(t)}|n(t)\ra;  \,\,  |\widehat{\psi}_n(t)\ra:=e^{i\beta^*_n(t)}|\widehat{n}(t)\ra. 
\eeqa
%
%(adiabaticity be particularized for each 
%mode as well). 
However, the  set of states $\{|\psi_n(t)\ra\}$ and the corresponding biorthogonal partners may be used to expand  
an arbitrary state, irrespective of its adiabaticity, as
\beq
\label{ansatz_beta}
|\Psi(t)\ra= \sum_n g_n(t)|\psi_n(t)\ra=\sum_n g_n(t) e^{i\beta_n} |n(t)\ra. 
\eeq

{\it Remark 1}: 
While our definition of adiabatic dynamics and the phases in Eq. ({\ref{beta}}) are  quite natural,
as they follow from the wave function ansatz and the Schr\"odinger equation, an alternative definition and phase  based 
instead on a fidelity criterion have been also proposed \cite{Comparat,qac}. The results are not always 
equivalent \cite{Comparat}. We restrict the present work to the definition
given above.  

{\it Remark 2}: Eq. (\ref{ansatz_betaad}) priviledges the time zero. 
There are often  physical reasons to do so, in particular when  
the eigenvectors of $H(0)$ at the preparation time $t=0$ form an orthonormal basis. 
A more general view is to 
associate adiabaticity of a mode $n$ with the approximate invariance of $|g_n(t)|^2$
in Eq. (\ref{ansatz_beta})  during some time  interval, that may or may not include the initial time.   
%where the constant might not be the initial value at $t=0$.     
%
%
%
%
%
\section{Generalized populations for the eigenstates of a non-Hermitian Hamiltonian}
The population of an instantaneous eigenvector $|n(t)\ra$ of a Hermitian Hamiltonian $H(t)$,
$P_n(t)=|\la n(t)|\Psi(t)\ra|^2$, 
may be formally generalized in many different ways for a NH system.  
Here are some possibilities (we shall frequently omit the time argument $t$ to avoid an overburdened notation):  
%\red{Here we propose some possibilities} (we shall frequently omit the time argument to avoid an overburdened notation):  
%
\beqa\label{genpro}
P_{1,n}&=&|\la \widehat{n}|\Psi\ra|^2  {= |c_n|^2},
\nonumber\\
\nonumber
P_{2,n}&=&\frac{|\la \Psi|\widehat{n}\ra \la n|\Psi\ra|}{\sum_n  |\la \Psi|\widehat{n}\ra \la n|\Psi\ra|},
% ||\Psi(t)||^2,
%\nonumber \\
%&=&  \frac{| c_n^* \sum_m \la n|m\ra c_m|}{\sum_n | c_n^* \sum_m \la n|m\ra c_m|} ||\Psi(t)||^2
\\\nonumber
P_{3,n}&=& \la \Psi|\widehat{n} \ra  \la n|n \ra  \la \widehat{n}|\Psi \ra,
% = |c_n|^2  \la n|n \ra,
\\\nonumber
P_{4,n}&=& \frac{\la \Psi| \widehat{n} \ra  \la \widehat{n}|\Psi \ra}{\la \widehat{n}|\widehat{n} \ra \la \Psi|\Psi \ra}, 
\\
P_{5,n}&=&|\la \widehat{\psi}_n|\Psi\ra|^2  {= |g_n|^2}. 
\eeqa
Their properties are summarized in Table I. 
They all tend to $P_n$ in the Hermitian limit, when $|\widehat{n}\ra=|n\ra$ become  orthonormal vectors. 
The list in Eq. (\ref{genpro})  is not exhaustive. For example, the roles of $|n\ra$ and $|\widehat{n}\ra$ may be reversed. We could even consider complex 
(instead of real) forms.
Also, some of them add up to one but, since the state norm  may  change in time for a NH system, it is natural to 
multiply the generalized populations by the square of the ordinary norm of the state,  $||\Psi||^2 =\la\Psi|\Psi\ra $, so that they sum up to $||\Psi||^2$.  
%In general they do not necessarily obey simple properties of ``probabilities'' such as $\sum_n P_n=1$, 
The $P_{j,n}$  do not necessarily obey the simple properties of proper  populations, such as
$\sum_n P_{j,n}=1$ and
$0\le P_{j,n}\le1$.
Some are $f$-dependent (they change with the change of basis $|n\ra\to|\phi_n\ra$ and  $|\widehat{n}\ra\to
|\widehat{\phi}_n\ra$), see Eqs. (\ref{set_eigenstates1}) and (\ref{set_eigenstates2}),
and others are not. 
The usefulness of these formal definitions will be determined by their physical content and the intended application. 
In particular,  since our main concern here is the characterization of adiabaticity, the property we should pay attention to
is ``adiabatic invariance''. An adiabatic invariant quantity remains constant when the state 
evolves according to Eq. (\ref{ansatz_betaad}).
%\red{This all is summarized in TABLE \ref{table}.}
%
\\
\begin{table}[h]
\begin{center}
    \begin{tabular}{ | c | c | c | c | c |}
    \hline
    $j$ & $\sum_n\! P_{j,n}\!=\!1$& $P_{j,n}\le 1$& $f$-indep. & adiab. inv. 
    \\ \hline
    1 & no & no & no &no 
    \\ \hline
    2 & yes & yes & yes &no  
    \\ \hline
    3 & no & no & yes& no
    \\ \hline
     4 & no & yes & yes & no
    \\ \hline
    5 & no & no & no & yes
    \\ \hline
    \end{tabular}
    \caption{Properties of different ``generalized populations'', see Eq. (\ref{genpro}).}
\end{center}
\label{table1}
\end{table}
The only definition in the group above which is adiabatically invariant independently of the reference basis 
 $\{|n(t)\ra\}$ chosen is $P_{5,n}=|g_n|^2$, so we shall examine  
its properties more carefully.  
The adiabatic invariance of $|g_n|^2$ is guaranteed by construction, but the values of the $g_n$ for an adiabatic evolution, however, 
will depend on the basis or, in other words,  be $f$-dependent, in a ``mild way''. 
Instead of  Eq. (\ref{ansatz_beta}), we can write the state of the system using a new basis as
\beq
\label{ansatz_beta1}
|\Psi(t)\ra\!=\!\! \sum_n \widetilde{g}_n(t) e^{-\frac{i}{\hbar}\!\int_0^t\! E_n(t') dt' - \int_0^t \la \widehat{\phi}_n(t')|\dot{\phi}_n(t')\ra dt'}
|\phi_n(t)\ra,
\eeq
where $\widetilde{g}_n(t)=\la\widehat{\phi}_n(t)|\Psi(t)\ra$. 
From Eq.  (\ref{cambio}),  
%$ \la \widehat{\phi}_{n}|\dot{\phi}_{n}\ra =  \dot{f}_n/f_n+ \la \widehat{n}|\dot{n}\ra$ 
taking into account Eqs. (\ref{set_eigenstates1}) and (\ref{set_eigenstates2}), it follows that
$e^{- \int_0^t \la \widehat{\phi}_{n}(t')|\dot{\phi}_{n}(t')\ra dt'} = e^{- \int_0^t \la \widehat{n}(t')|\dot{n}(t')\ra dt'} f_n(0)/f_n(t)$,
since $e^{- \int_0^t \dot{f}_n(t')/f_n(t') dt'} = e^{\ln[f_n(0)]-\ln[f_n(t)]}$.
Thus, comparing  Eqs. (\ref{ansatz_beta1}) and  (\ref{ansatz_beta}), we find that  
$\widetilde{g}_n(t) = g_n(t)/f_n(0)$.  The only difference between the values {of these amplitudes} for different bases,  
independently of adiabaticity, is a constant factor, {$f_n(0)$,} that depends on the initial normalization.   
In many processes of physical interest, the natural basis at $t=0$ is orthogonal. This does not
necessarily imply that $H(0)$ is Hermitian. For example, in the Landau-Zener or 
coherent population return processes that we shall discuss in Secs. IV and V
for a two-level system,
spontaneous decay is always present, even at $t=0$, so $H(t)$ is never Hermitian. However, before switching the
laser on, the ``bare'' basis  formed by atomic  ground and excited states is orthogonal,
{$\la n(0)|n'(0)\ra=0$, if $n\ne n'$}.
In principle, it would still be possible to distinguish $|n(0)\ra$ 
and $|\widehat{n}(0)\ra=|n(0)\ra/\la n(0)|n(0)\ra$,
{from Eq. (\ref{delta})},
but the simplest and most useful 
convention is to set $\la n(0)|n(0)\ra=1$, so that $|\widehat{n}(0)\ra=|n(0)\ra$ and 
$\{|n(0)\ra\}$ becomes an ordinary orthonormalized basis at the initial instant of time.    
Then the ${P_{5, n}(0)=} |g_n(0)|^2$ become ordinary populations, ${P_n(0)}$.
Hereafter we shall limit the discussion to this 
{type of systems and convention}. 
{From Eqs. (\ref{set_eigenstates1}) and (\ref{set_eigenstates2}),} the only allowed $f_n(0)$ to satisfy the orthonormalization {condition, $\la n(0)|n'(0)\ra=\delta_{n, n'}$,} are of modulus one, 
so $|\widetilde{g}_n(t)|^2=|g_n(t)|^2$, 
even if the system does not follow adiabatic dynamics.
For non-adiabatic dynamics, the $|g_n(t)|^2$ are not bounded by one, and their sum 
over $n$ may  be anything. However,  if  $ \sum_n |g_n(0)|^2=1$, the sum will still be one as long as the evolution remains adiabatic for all states,  $\sum_n |g_n(t)|^2=1$. 
Thus, $``1"$ becomes  the relevant scale to identify adiabaticity or its absence.
For a state that begins 
like $|\Psi(0)\ra=|m(0)\ra$,  with $|g_m(0)|^2=1$, adiabatic dynamics implies $|g_m(t)|^2\approx1$, whereas 
for $n\ne m$, $|g_n(0)|^2=0$, adiabaticity implies $|g_n(t)|^2\ll 1$.

We may consider instead of  Eqs. (\ref{ansatz_beta}) and (\ref{ansatz_beta1}) the expansions 
\beqa
|\Psi(t)\ra &=& \sum_n d_n(t) e^{- \int_0^t \la \widehat{n}(t')| \dot{n}(t') \ra dt'} |n(t)\ra,
\\ 
|\Psi(t)\ra &=& \sum_n \widetilde{d}_n(t) e^{- \int_0^t \la  \widehat{\phi}_n(t')| \dot{\phi}_n(t') \ra dt'}  |\phi_n(t)\ra,
\eeqa
see \cite{jolicard}, without an explicit dynamical factor. ({If the basis is parallel transported $d_n(t)=c_n(t)$.}) The coefficients $d_n(t)$ are also weakly dependent on a basis change, i.e., they obey 
$\widetilde{d}_n(t)=d_n(t)/f_n(0)$. However, for NH systems they may suffer strong, 
exponential variations even for adiabatic dynamics, as 
 \beq
\label{gn}
d_{n}(t) = g_n(t) e^{\frac{-i}{\hbar}\int_0^t E_n(t') dt'},
\eeq
and the $E_n$ are generally complex.  
As a consequence, the ratios 
$|d_{n}(t)|^2/|d_{n'}(t)|^2$, {for $n \neq n'$}, change dramatically due to different exponential 
dynamical factors even when the  two implied states behave  adiabatically.
In any case the $d_n$ coefficients may be physically very relevant. If a  parallel-transported basis
$\{|n(t)\ra\}$  becomes orthonormalized again at the final process time $t_f$, the $|d_n(t_f)|^2$ would directly give actual 
populations, unlike the  $|g_n(t_f)|^2$, generally affected by suppressing or enhancing dynamical exponentials.  This is important because 
non-adiabatic excitations revealed  in the $\{|\psi_n\ra\}$ basis by a large $|g_n(t_f)|^2$ value might actually 
be irrelevant in practice  if the corresponding $|d_n(t_f)|^2$ turns out to be negligible.
In general it is advisable to analyze a given process simultaneously in different bases.        
{\section{Approximate adiabaticity condition for non-Hermitian Hamiltonians}}
%
%
%
%
%\subsection{Introduction}
%
%
%
%
The standard (Hermitian) adiabaticity criterion is only valid in the weak-non-Hermiticity regime \cite{nenciu}, in  which the absolute values of the imaginary parts of the eigenvalues are smaller or of the same order as the slowness parameter. In this case a  generalization of the adiabatic theorem for Hermitian Hamiltonians can be done
for non-degenerate eigenvalues 
%(when degeneracies exist the situation is more subtle)
\cite{putterman, wright}. 
Instead, in the ``strong-non-Hermiticity regime" at least some of the eigenvalues have imaginary parts with absolute values much larger than the slowness parameter. In this case a complete generalization of the adiabatic theorem is not possible but an adiabatic  theorem-like result can be worked out for the least dissipative eigenvalue \cite{nenciu}. 
%, if it remains so throughout the process.  

As in \cite{sar_sof} and \cite{erratum},
assuming that the general state of the system is given by Eq. (\ref{ansatz_beta}), parallel transported eigenstates so that 
$\la\widehat{n}|\dot{n}\ra = 0$,
and inserting Eq. (\ref{ansatz_beta}) into the Schr\"{o}dinger equation (\ref{Schr_eq}),  we get 
\beqa
\label{approx_cond0}
\dot{g}_n(t)&=&
-\sum_{k\neq n} e^{iW_{nk}(t)} \la\widehat{n}(t)|\dot{k}(t)\ra g_k(t),
%\sum_{k\neq n}\frac{g_k(t)}{\hbar \omega_{nk}(t)}\left[\exp\left(i\!\int_0^t\! \omega_{nk}(t') dt'\right)\right]
%\\\nonumber
%&\times&\left\la \widehat{\phi}_n(t)\left|\dot{H}_0\right|\phi_k(t)\right\ra,  
\eeqa
where $W_{nk}(t)=\int_0^t\! \omega_{nk}(t') dt'$ and   $ \omega_{nk}(t):=[E_n(t)-E_k(t)]/\hbar$. 
%, and
%the sum $\sum^{'}$ is for $k\neq n$. 
%$\la \widehat{\phi}_n(t)|\dot{H}_0|\phi_k(t)\ra=[E_k(t)-E_n(t)] \la \widehat{\phi}_n(t)|\dot{\phi}_k(t)\ra$.
Integrating this formally gives  
\beq
\label{approx_cond}
g_n(t)-g_n(0) = -\sum_{k\neq n} \int_0^t e^{iW_{nk}(t')} \la\widehat{n}(t')|\dot{k}(t')\ra g_k(t') dt' .
\nonumber
\eeq
We now apply perturbation theory. 
Assuming that the system is initially in  $|m(0)\ra$,
and approximating the coefficients $ g_k(t)$ inside the integral 
as $g_k(t)=\delta_{km}$,
one finds in  first order, for $n\ne m$, 
%Assuming that the system is initially at a state $|\phi_m(0)\ra$, $g_m(t)=1$, we have in first order \red{approximation},
%for $n\neq m$,   
%
\beq
\label{aco}
g_n(t) =  -\int_0^t  \la \widehat{n}(t')|\dot{m}(t')\ra e^{iW_{nm}(t')} dt',
\eeq
% 
%where $W_{nm}(t')=\int_0^{t'}\! \omega_{nm}(t'') dt''$.
{which should satisfy $|g_n(t)| \ll 1$ for an adiabatic evolution.}
Rewriting Eq. (\ref{aco}) 
as $g_n(t)=-\int_0^t u_n dv_n$,
with 
\beqa
u_n&=&\frac{\la \widehat{n}(t')|\dot{m}(t')\ra}{i\omega_{nm}(t')},
\nonumber\\ 
dv_n&=&i\omega_{nm}(t')e^{iW_{nm}(t')} dt',
\label{uv}
\eeqa  
and integrating by parts, 
we find
\beq
\label{parts}
g_n(t)=-\frac{\la \widehat{n}(t')|\dot{m}(t')\ra}{i\omega_{nm}(t')}e^{iW_{nm}(t')}\Bigg|_0^t
+\int_0^t v_n du_n.
%\frac{e^{iW_{nm}(t')}[\dot{\omega}_{nm}(t')\la \widehat{\phi}_n(t')|\dot{\phi}_m(t')\ra-\omega\frac{d<>}{dt}}{i%\omega_{nm}^2(t')} dt'
%\nonumber
\eeq
Neglecting the integral term {in Eq. (\ref{parts})}, which, as shown in  Appendix A,  involves higher inverse powers of $\omega_{nm}$, and the 
(generally small) contribution at $t=0$,  we get from $|g_n(t)|\ll 1$ the approximate adiabaticity condition
\beq
\label{uv_n}
|(uv)_n(t)| = \frac{|\la \widehat{n}(t)|\dot{m}(t)\ra|}{|\omega_{nm}(t)|}e^{- {\rm Im}[W_{nm}(t)]}  \ll 1,
\eeq
for $n \neq m$.
%, where $\rm Im$ means the imaginary part. 
For $n=m$ a second order integral may be written but it does not lead to
a simple expression by integration by parts. The fact that the condition (\ref{uv_n}) is limited to $n\neq m$ is quite harmless for Hermitian systems, 
because of the conservation of total probability and the orthogonality of states. 
In a NH system it is a more serious limitation, as we cannot deduce 
from it the adiabaticity or otherwise 
of the initially occupied state. 
The criterion (\ref{uv_n}) is a natural generalization of the usual Hermitian criterion, and it  outperforms  
other approximations based on partitions of Eq. (\ref{aco}) alternative to Eq. (\ref{uv}), such as \cite{ChangPu}
\beqa
u'_n&=&\frac{\la \widehat{n}(t')|\dot{m}(t')\ra e^{-{\rm Im}{[W_{nm}(t')]}} }{i{\rm Re}[\omega_{nm}(t')]},
\nonumber\\
\label{uv'}
dv'_n&=&i{\rm Re}[\omega_{nm}(t')]e^{i{\rm Re}[W_{nm}(t')]} dt'.
\eeqa
%
%where $\rm Re$ means the real part. 
Similarly we could try  
\beqa
u''_n&=&\frac{\la \widehat{n}(t')|\dot{m}(t')\ra e^{i{\rm Re}[{W_{nm}(t')]}} }{-{\rm Im}[\omega_{nm}(t')]},
\nonumber\\\label{uv''}
dv''_n&=&-{\rm Im}[\omega_{nm}(t')]e^{-{\rm Im}[W_{nm}(t')]} dt'.
\eeqa  
%
%which fails at processes such as the one depicted in Fig. ...
These partitions lead to conditions similar to Eq. (\ref{uv_n}), but with $|\text{Re}(\omega_{nm})|$,
for Eq. (\ref{uv'}),
and $|\text{Im}(\omega_{nm})|$, for Eq. (\ref{uv''}), in the denominator. In processes such as a Landau-Zener transition 
for a two-level atom
discussed later, the real or the imaginary parts of the energies may become equal
for some $t$, but $|\omega_{nm}|$ is always different from
zero as long as fully degenerate points (with equal eigenvalues) are not crossed.
%
%{\rm{Im}}[\omega_{nm}] instead of {\rm{Re}}[\omega_{nm}]. 
%
%
%
%
%
%
%

$^{}$
\section{\label{model}Model: Landau-Zener and Coherent Population Return processes for a two-level atom}
We shall exemplify the previous analysis with two types of adiabatic processes of physical interest for a decaying two-level atom:
a Landau-Zener (LZ) protocol with constant laser intensity, which in the  appropriate parameter range
produces population inversion;      
% illuminated by a chirped laser pulse, i.e, one with a time-dependent frequency passing through resonance, 
and coherent population return (CPR) with constant laser detuning and Gaussian Rabi frequency, a useful process to suppress power broadening \cite{CPR}. 
We assume  for simplicity that a Hamiltonian description,  rather than a master equation,  is enough for the trapped atom \cite{sar_sof, Pritchard,Lizuain}. This happens, for example, when the decayed atom escapes from the trap by recoil. 
We shall also assume a semiclassical treatment of the interaction between the 
electric field and the atom, as well as a constant decay rate $\Gamma$, the inverse life-time, from the excited state to the ground state. 
%The decay rate could in general be time dependent, $\Gamma(t)$.

Applying the electric dipole approximation, a laser-adapted interaction picture, and the rotating wave approximation, 
the Hamiltonian, disregarding atomic motion, is \cite{sara}
\beq
H_{a0}(t)=\frac{\hbar}{2}
\left(\begin{array}{cc}
-\Delta(t) & \Omega_{R}(t)\\
\Omega_{R}(t) & \Delta(t)-i\Gamma
\end{array} \right),
\label{hami}
\eeq
in the bare basis $|{\sf g}\rangle = \left( \begin{array} {rccl} 1\\ 0 \end{array} \right)$ and $|{\sf e}\rangle = \left( \begin{array} {rccl} 0\\ 1 \end{array} \right)$ of the atom.
The norm of the general state $|\Psi(t)\ra$ decreases due to spontaneous decay. 
%since, if an atom decays spontaneously from level $2$ to level $1$, it is eliminated from the quantum ensemble.  
The detuning is defined as 
$\Delta(t)=\omega_0-\omega(t)$, where $\omega(t)/(2\pi)$ is the instantaneous field frequency
and $\omega_0/(2\pi)$ the transition frequency. 
The (real) Rabi frequency $\Omega_R(t)$ in general also depends on time.
%We consider a pulse with slowly varying envelope so that 
%the (real) Rabi frequency
%$\Omega_R(t)$ also depends on time.
The eigenvalues of this Hamiltonian are
\beq
\label{E}
E_{\pm}(t) = \frac{\hbar}{4} \left\{-i\Gamma \pm \sqrt{-[\Gamma+2i\Delta(t)]^2 + 4\Omega^2_R(t)} \right\},
\eeq
and the right eigenstates,  that play the role of $\{|n(t)\ra\}$ here,   are
\beqa
\nonumber
|{+}(t)\ra = \sin\left(\frac{\alpha}{2}\right) |{\sf g}\ra + \cos\left(\frac{\alpha}{2}\right) |{\sf e}\ra,
\\
\label{n-}
|{-}(t)\ra = \cos\left(\frac{\alpha}{2}\right) |{\sf g}\ra - \sin\left(\frac{\alpha}{2}\right) |{\sf e}\ra,
\eeqa
where the mixing angle $\alpha=\alpha(t)$ is complex and defined from
\beq
\label{alpha}
\tan [\alpha(t)] = \frac{\Omega_R(t)}{\Delta(t) - i \Gamma /2}
\eeq
as $\alpha= \arctan(x) = i [\ln(1-i x)-\ln(1+i x)] /2$,   with 
\beq
\label{x}
x(t)=  \frac{\Omega_R(t)} {\Delta(t) - i \Gamma /2}.
\eeq

The adjoint of $H_{a0}(t)$ is
\beq
H_{a0}^{\dag}(t)=\frac{\hbar}{2}
\left(\begin{array}{cc}
-\Delta(t) & \Omega_{R}(t)\\
\Omega_{R}(t) & \Delta(t)+i\Gamma
\end{array} \right),
\eeq
with eigenvalues $E_{\pm}^*(t)$ and right eigenstates
\beqa
\nonumber
|\widehat{+}(t)\ra = \sin\left(\frac{\alpha^*}{2}\right) |{\sf g}\ra + \cos\left(\frac{\alpha^*}{2}\right) |{\sf e}\ra,
\\
\label{n-1}
|\widehat{-}(t)\ra= \cos\left(\frac{\alpha^*}{2}\right) |{\sf g}\ra - \sin\left(\frac{\alpha^*}{2}\right) |{\sf e}\ra.
\eeqa
The coefficients are complex conjugate of those in 
Eq. (\ref{n-}) because 
$H_{a0}(t)$ is equal to its transpose \cite{Muga}.
For later use we calculate  the matrix elements 
\beqa
\label{paral_trans}
\la\widehat{\pm}(t)|\partial_{t} {\pm}(t)\ra &=& 0,
\\
\la\widehat{\mp}(t)|\partial_{t} {\pm}(t)\ra &=& \pm \frac{\dot{\alpha}}{2} ,
\eeqa
where
\beqa
\dot{\alpha}(t) = \frac{\dot{\Omega}_R(t) [\Delta(t) - i \Gamma/2]- \Omega_R(t) \dot{\Delta}(t) }
{[\Delta(t)-i \Gamma/2]^2 + \Omega^2_R(t)}.
\eeqa
Eq. (\ref{paral_trans}) shows that the states $|+(t)\ra$ and $|-(t)\ra$ are parallel transported.
To impose the continuity of the eigenvalues and eigenvectors throughout the process and the correct matching 
of their $\pm$ labels we have to choose the branches 
of the square root in Eq. (\ref{E}) and of the $\arctan{(x)}$. 
%Here, we are not identifying the eigenvalues with the eigenstates, note that the subindex in Eqs. (\ref{E}) and (\ref{n-}) do not coincide, since 
For each protocol $\Delta(t)$ and $\Omega_R(t)$ are specified and we have to analyze 
the behaviour of the radicand in Eq. (\ref{E}),
\beq
\label{z}
z(t) = -[\Gamma+2i\Delta(t)]^2 + 4\Omega^2_R(t). 
\eeq
%
%is different making necessary different elections of the branches of $\sqrt{z(t)}$. 
%Consider first the branch cut right below the negative real axis. This 
%makes $\text{Re}(\sqrt{z}) > 0$. 
$z(t)$ is in polar form
$z= R e^{i \gamma}$, with modulus 
$R= |\sqrt{\text{Re}(z)^2 + \text{Im}(z)^2}|$  and argument $\gamma$, where  
$\text{Re}(z)=4\Delta^2(t) +4\Omega_R^2(t)-\Gamma^2$ and $\text{Im}(z)=-4\Delta(t) \Gamma$.

The first process we consider is the Landau-Zener protocol, 
with linear detuning and constant Rabi frequency, 
\beqa
\label{delta_lz}
\Delta_{LZ}(t) &=& b (t-t_f/2),
\\
\Omega_{R, LZ}(t) &=& \Omega_0,
\eeqa
where $t_f$ is the final time of the process, $b>0$ is the constant ``chirp", and $\Omega_0$ is the constant Rabi frequency.
%If the two-level system decays coherently from the upper level with slow spontaneous decay driven by a laser,
%the population of the excited state $|2\ra$ at the initial instant of time
%is $P_2(0)=1$, whereas the population of the ground state is $P_1(0)=0$.
%The coherent decay may be driven adiabatically
%with a ``rapid'' adiabatic passage (RAP) technique \cite{NMR}, sweeping the laser frequency across
%resonance. The adjective ``rapid'' here could be misleading: it simply means ``faster
%than the spontaneous decay'' but, as the approach is 
%based on an adiabatic passage,
%it fails for short enough pulse times. 
%
Two  regimes can be distinguished for this protocol depending on 
whether $\Gamma < 2 \Omega_0$ or $\Gamma > 2 
\Omega_0$
{(a degeneracy exists  at $t=t_f/2$ if 
$\Gamma = 2 \Omega_0$):}

{\it{(i)}} When $\Gamma < 2 \Omega_0$, then {$\text{Re}(z)>0$. 
%
%\beq
%\label{Rez}
%\text{Re}(z) = .
%\eeq
%
A representative trajectory of $z$ {in the complex $z$ plane
% for the parameters
%$\Gamma =2\pi\times0.159$ MHz, $\Omega_0 =2\pi\times0.159$ MHz,
%$b =2\times 10^{12}$ s$^{-2}$, and $t_f=3$ $\mu$s} 
is shown in 
%of $z$ 
%for this case 
Fig. \ref{fig_z,x} (a). 
We choose the branch cut of the square root just below the negative real axis, so that $-\pi<\gamma\le\pi$. 
%
%%%%%%%%%%%%%%%%%%%%%%%%%%%%%%%%%%%%%%%%%%%%%begin figure%%%%%%%%%%%%%%%%%%%%%%%%%%%%%%%%%%%%%%%%%%%%%%%%%%%%%%%%%
\begin{figure}[t]
\begin{center}
\includegraphics[height=2.1cm,angle=0]{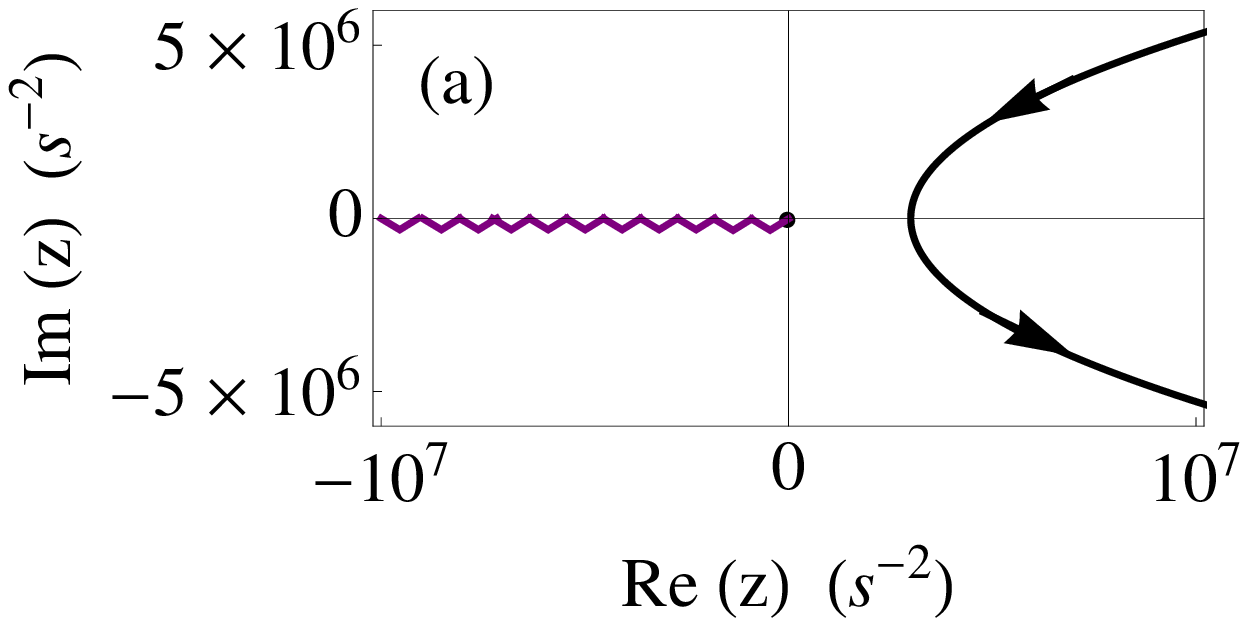}
\includegraphics[height=2.1cm,angle=0]{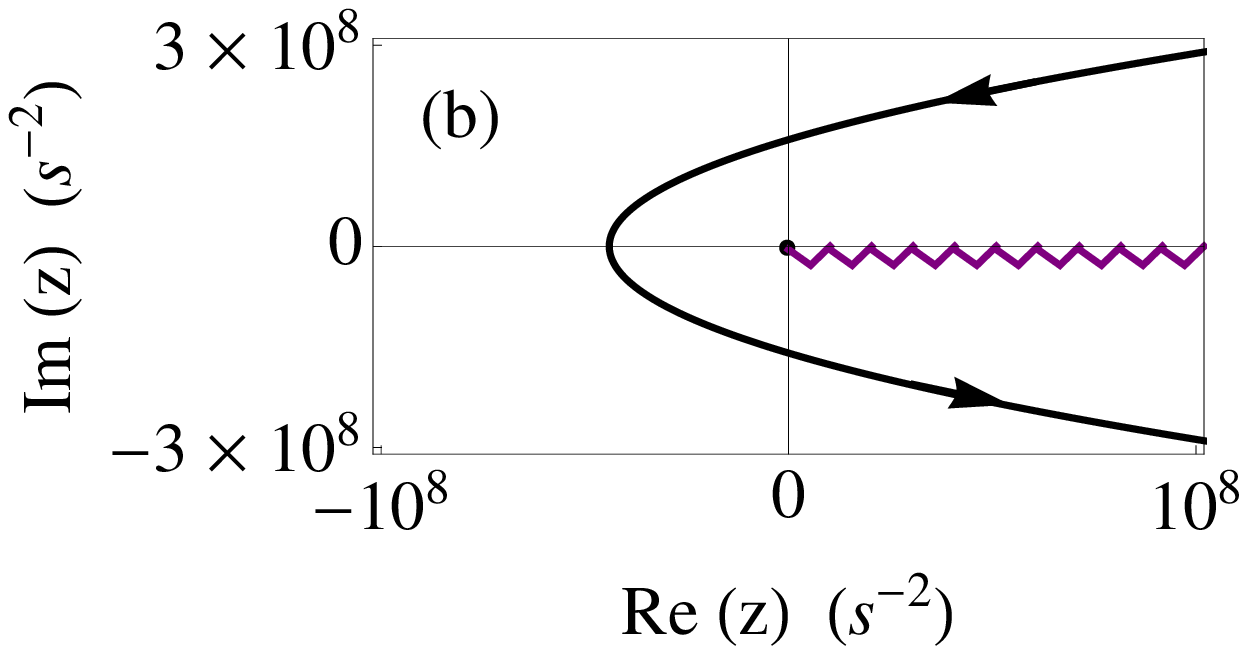}
\includegraphics[height=2.1cm,angle=0]{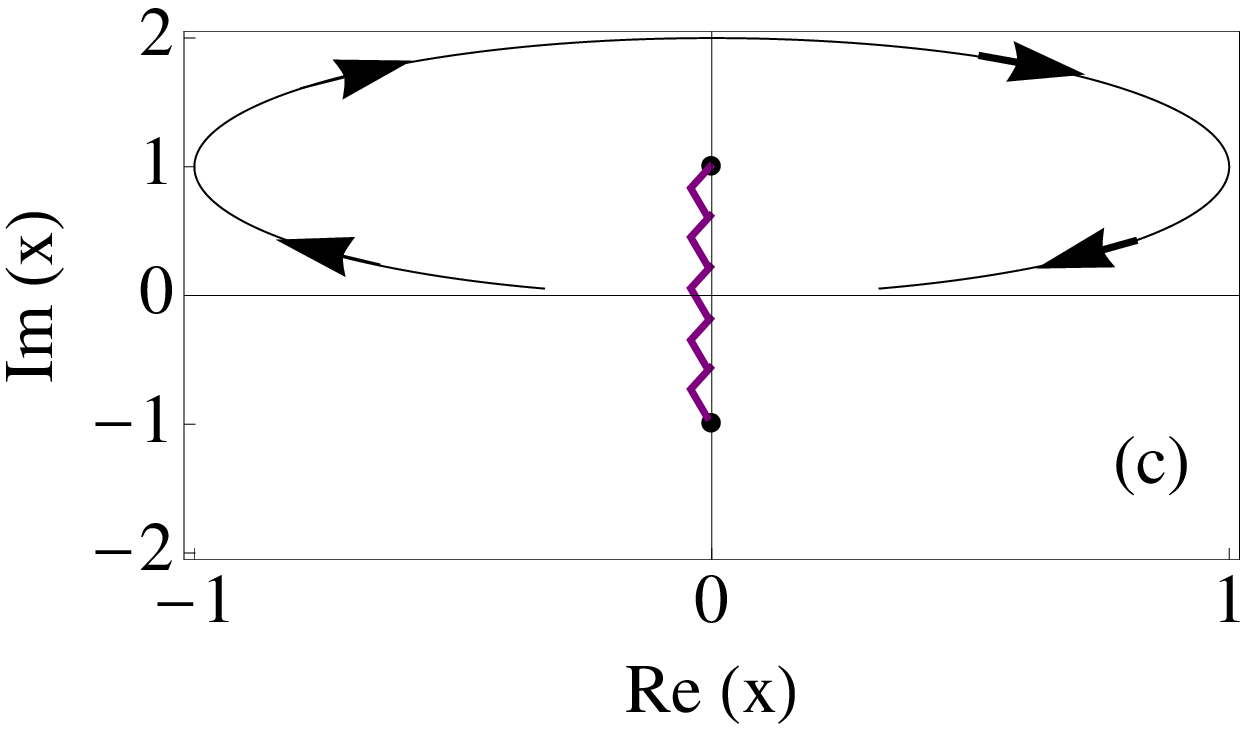}
\includegraphics[height=2.1cm,angle=0]{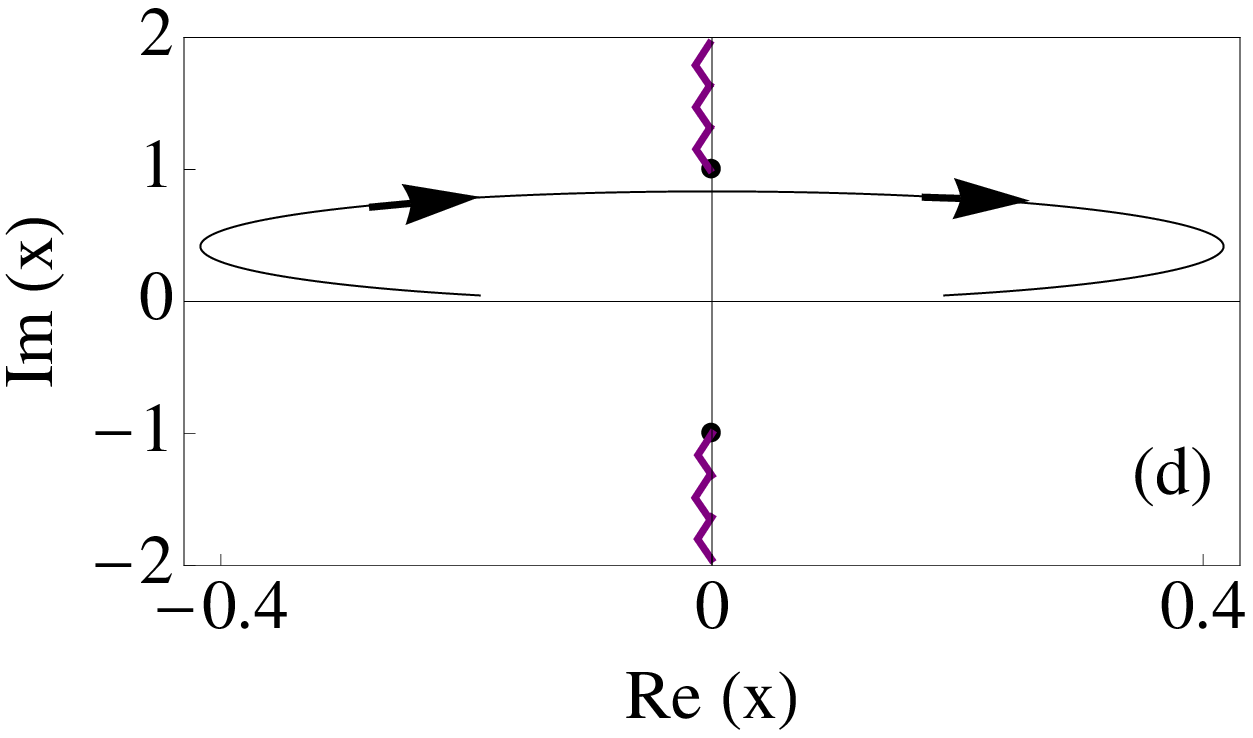}
\end{center}
\caption{\label{fig_z,x}
(Color online) 
Landau-Zener processes.
Branch cuts and representative trajectories in the complex $z$ and $x$ planes, see Eqs. (\ref{z}) and (\ref{x}). 
(a,c): 
$\Gamma < 2 \Omega_0$  
($\Gamma =2\pi\times0.159$ kHz, $\Omega_0 =2\pi\times0.159$ kHz,
$b =2\times 10^{6}$ s$^{-2}$, and $t_f=3$ ms);  (b,d): $\Gamma > 2 \Omega_0$ ($\Gamma =2\pi\times1.910$ kHz, $\Omega_0 =2\pi\times0.796$ kHz, $b =50\times10^{6}$ s$^{-2}$, and $t_f=1$ ms).
In (a) the branch cut, just below the negative real axis, is chosen so that $-\pi<\gamma\le\pi$; in (b), just below the positive real axis, is chosen so that $0 \le \gamma <2 \pi$. %The branch cuts in (c) and (d) are chosen for continuity and in (d) we add $\pi$ to $x$ to match the $\pm$-labeling of eigenvectors and eigenvalues.
}
\end{figure}
%%%%%%%%%%%%%%%%%%%%%%%%%%%%%%%%%%%%%%%%%%%%%end figure%%%%%%%%%%%%%%%%%%%%%%%%%%%%%%%%%%%%%%%%%%%%%%%%%%%%
%Thus, there are not discontinuities for $\sqrt{z}$ and the energies are well defined by the expressions given in Eq. (\ref{E}). We rename them as $E_{\pm}(t) = \varepsilon_{\pm}(t)$.
The imaginary parts of both energies cross each other at $t=t_f/2$,
where $\text{Im}[E_+(t_f/2)]= \text{Im}[E_- (t_f/2)]= -i \hbar \Gamma/4$, and the real parts have an avoided crossing at this instant of time, 
see Figs. \ref{fig_E} (a) and \ref{fig_E} (b).
With this branch election $\text{Im}[E_+(t)] > \text{Im}[E_- (t)]$ when $t < t_f/2$ and 
$\text{Im}[E_-(t)] >  \text{Im}[E_+ (t)]$
when $t > t_f/2$, which implies that the least dissipative state changes from  
$|+(t)\ra$ when $t<t_f/2$ to  $|-(t)\ra$ when $t>t_f/2$.  
%
%%%%%%%%%%%%%%%%%%%%%%%%%%%%%%%%%%%%%%%%%%%%%begin figure%%%%%%%%%%%%%%%%%%%%%%%%%%%%%%%%%%%%%%%%%%%%%%%%%%%%%%%%%
\begin{figure}[t]
\begin{center}
\includegraphics[height=2.1cm,angle=0]{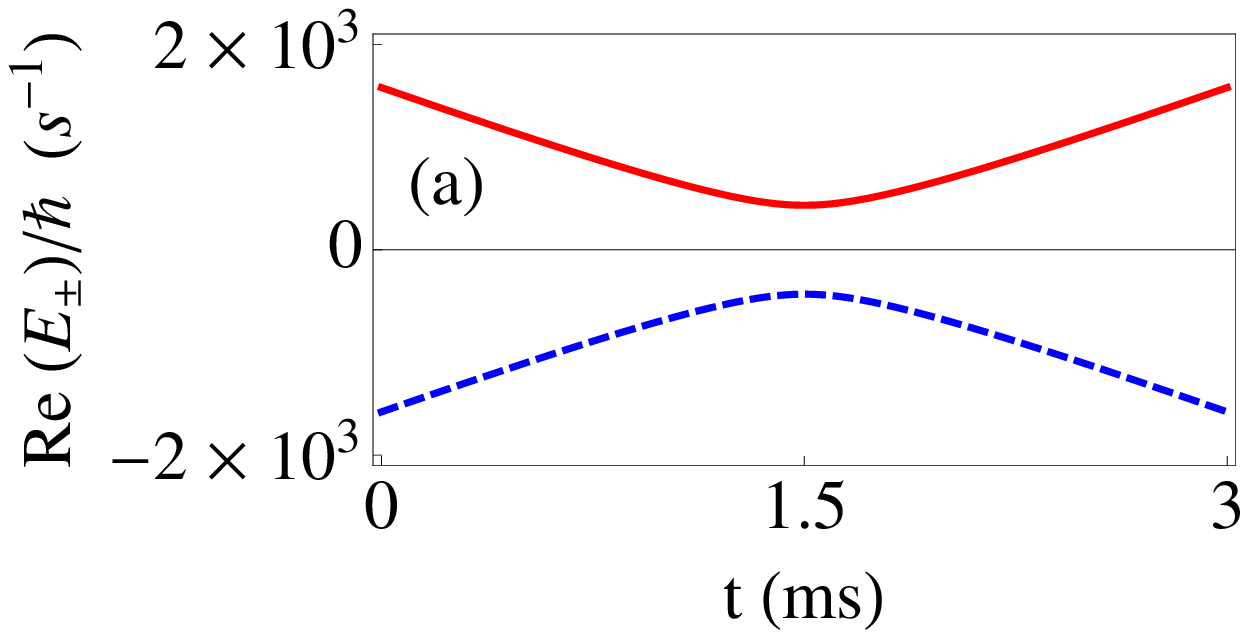}
\includegraphics[height=2.1cm,angle=0]{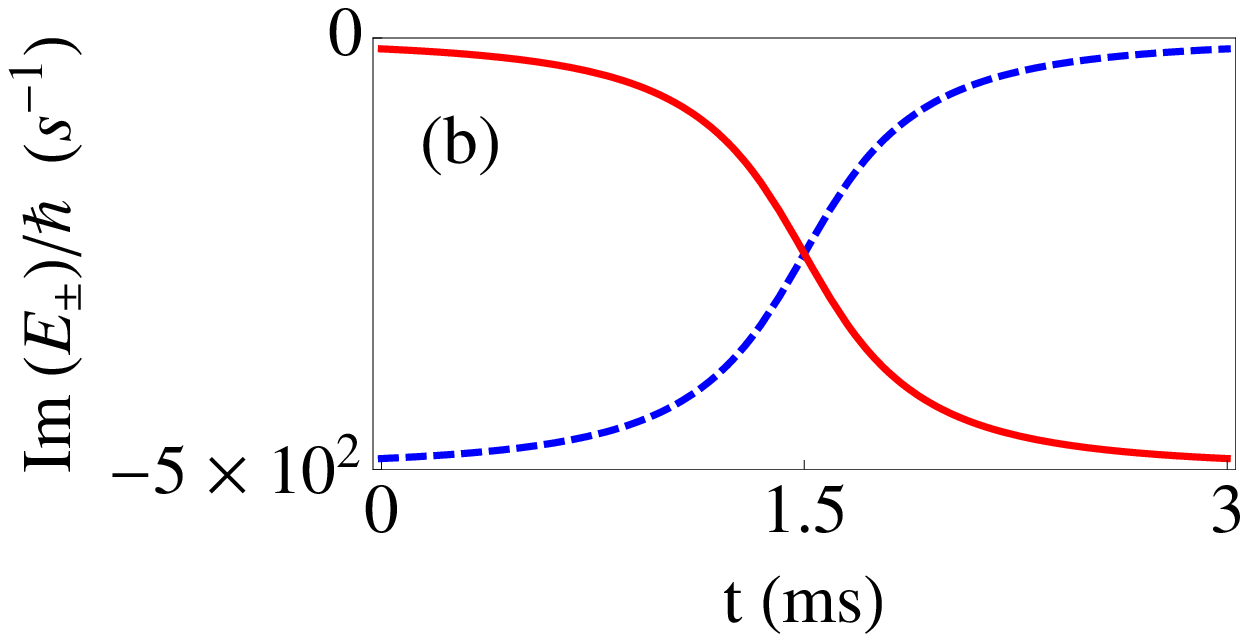}
\includegraphics[height=2.1cm,angle=0]{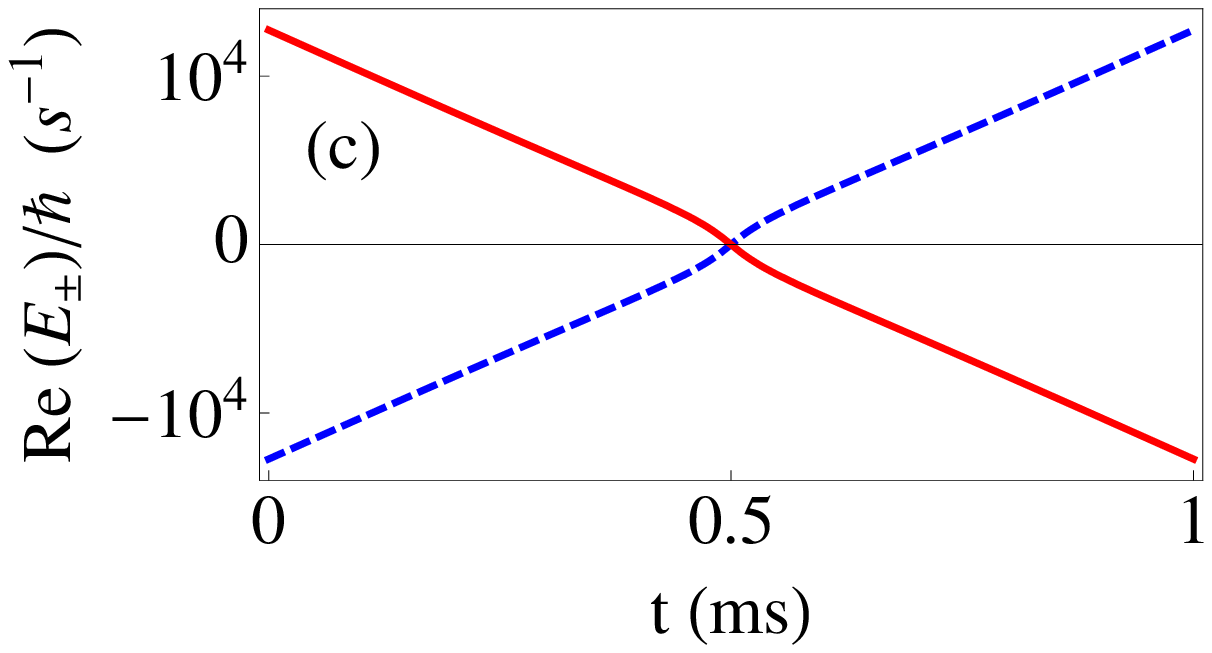}
\includegraphics[height=2.1cm,angle=0]{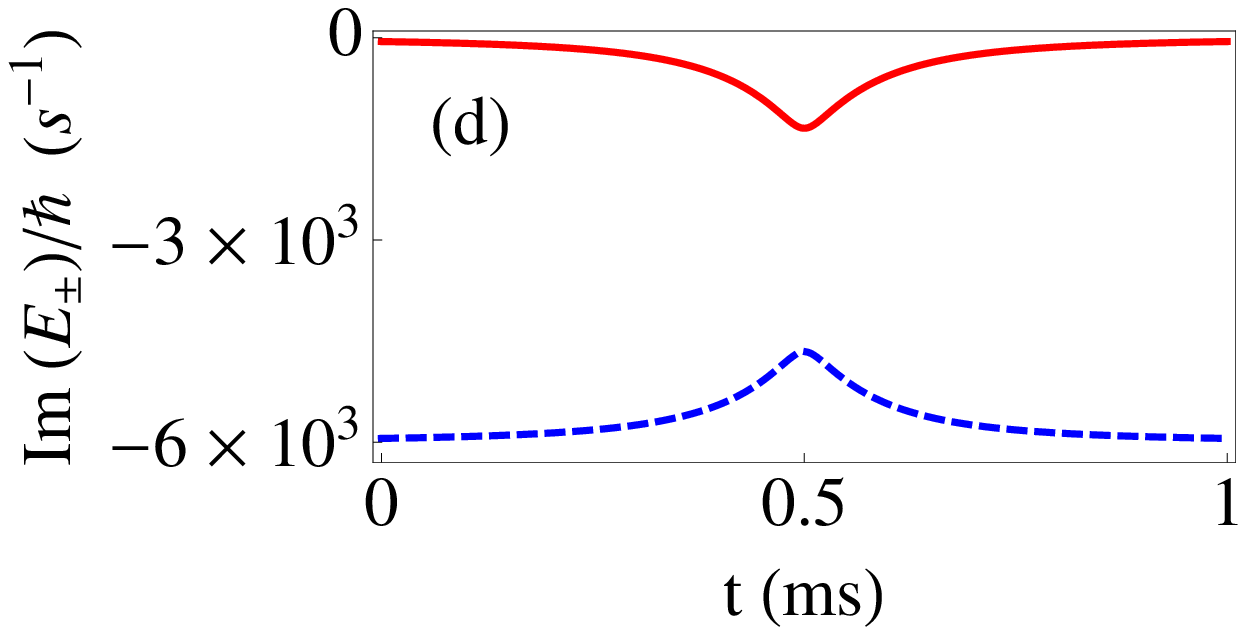}
\includegraphics[height=2.1cm,angle=0]{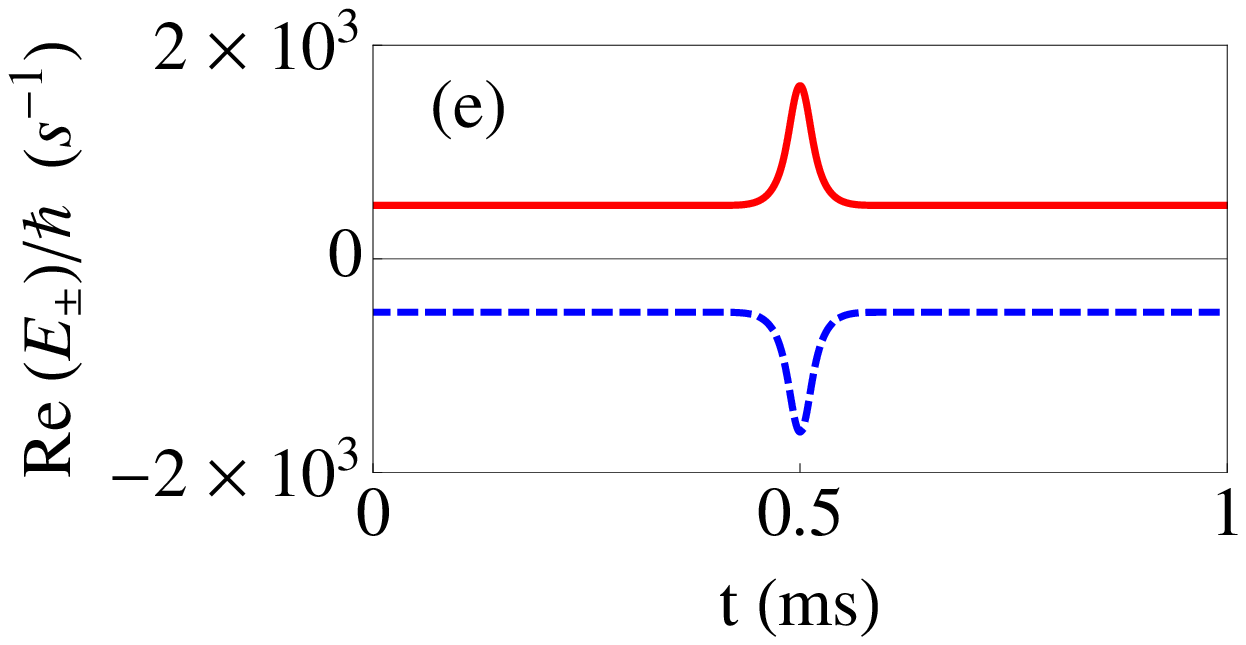}
\includegraphics[height=2.1cm,angle=0]{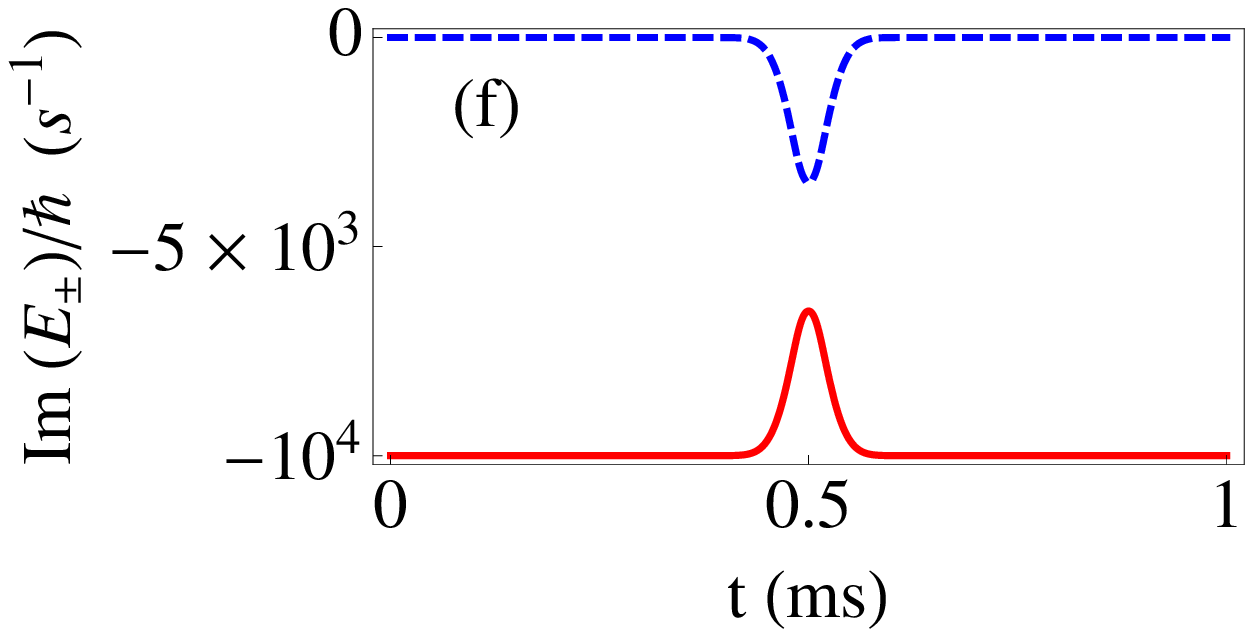}
\end{center}
\caption{\label{fig_E}
(Color online) Real and imaginary parts of the energies for a Landau-Zener and CPR processes. 
(a,b): LZ, 
$\Gamma < 2 \Omega_0$; (c,d): LZ, $\Gamma > 2 \Omega_0$; 
(e,f): CPR.  
$E_+$: red solid line; $E_-$: blue dashed line. 
Parameters as in Fig. \ref{fig_z,x} for LZ and $\Gamma =2\pi\times3.183$ kHz, $\Omega_{max} =2\pi\times1.592$ kHz, $a =4\times10^{8}$ s$^{-2}$, $\Delta_0=2\pi \times0.159$ kHz, and $t_f=1$ ms for CPR.
}
\end{figure}
%%%%%%%%%%%%%%%%%%%%%%%%%%%%%%%%%%%%%%%%%%%%%end figure%%%%%%%%%%%%%%%%%%%%%%%%%%%%%%%%%%%%%%%%%%%%%%%%%%%%
%
The initial detuning is negative, {see Eq. (\ref{delta_lz}),} and the trajectory of $x$ in the complex plane is depicted in
Fig. \ref{fig_z,x} (c),  so we choose for continuity 
the $\arctan{(x)}$ branch cut in that figure. Note the inversions $|+(0)\ra\approx |{\sf g}\ra\to |+(t_f)\ra\approx |{\sf e}\ra$,  
$|-(0)\ra\approx -|{\sf e}\ra\to |-(t_f)\ra\approx|{\sf g}\ra$ as $\alpha(0)\approx \pi \to\alpha(t_f)\approx 0$.  This model describes Rapid Adiabatic Passage (RAP) by a LZ protocol in presence of decay. 

{\it{(ii)}} When $\Gamma > 2 \Omega_0$, 
%from Eq. (\ref{Rez}), 
$z$ crosses the negative real axis  
as shown in Fig. \ref{fig_z,x} (b),  
%\red{for the set of parameters
%$\Gamma =2\pi\times1.910$ MHz, $\Omega_0 =2\pi\times0.796$ MHz, $b =50\times10^{12}$ s$^{-2}$, and $t_f=1$ $\mu$s,}
and we choose the branch cut for the square root just below the positive real axis, so that $0\le \gamma<2\pi$. 
% for parameters:
%
Now the real parts of $E_{\pm}(t)$ cross at $t=t_f/2$, where $\text{Re}[E_+(t_f/2)]= \text{Re}[E_-(t_f/2)]= 0$, and
$\text{Im}[E_+(t)] > \text{Im}[E_- (t)]$, see Figs. \ref{fig_E} (c) and \ref{fig_E} (d). Thus, $|+(t)\ra$ is the least dissipative state for the whole process.
The form of the $x$-trajectory is depicted in Fig. \ref{fig_z,x} (d).
We choose the branch cuts as depicted in the figure to assure
continuity
%\footnote{This is the branch cut by default in Mathematica.}
and add $\pi$ to define $\alpha$ so as to match the $\pm$-labeling of 
eigenvectors and eigenvalues. 
$\alpha$ evolves from $\alpha(0)\approx\pi$ to $\alpha(t_f)\approx\pi$
and the eigenvectors are not inverted: $|+(0)\ra\approx|+(t_f)\ra\approx|{\sf g}\ra$,  $|-(0)\ra\approx|-(t_f)\ra\approx-|{\sf e}\ra$.

%
%%%%%%%%%%%%%%%%%%%%%%%%%%%%%%%%%%%%%%%%%%%%%begin figure%%%%%%%%%%%%%%%%%%%%%%%%%%%%%%%%%%%%%%%%%%%%%%%%%%%%%%%%%
\begin{figure}[t]
\begin{center}
\includegraphics[height=4.0cm,angle=0]{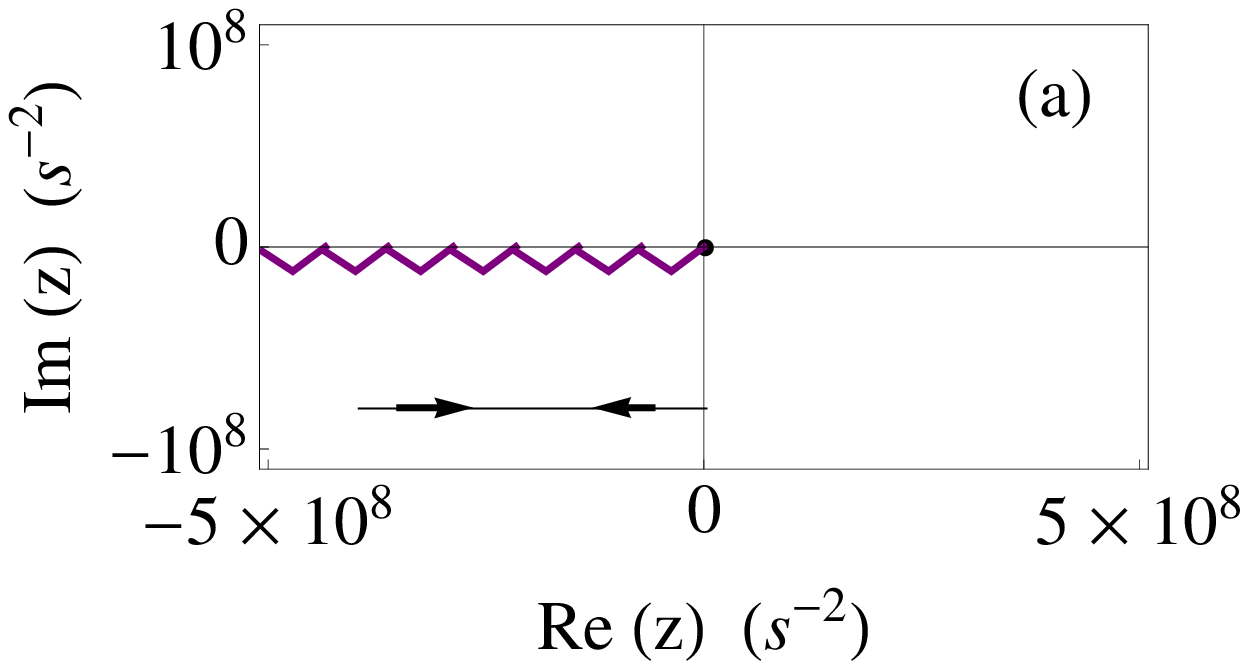}
\includegraphics[height=4.0cm,angle=0]{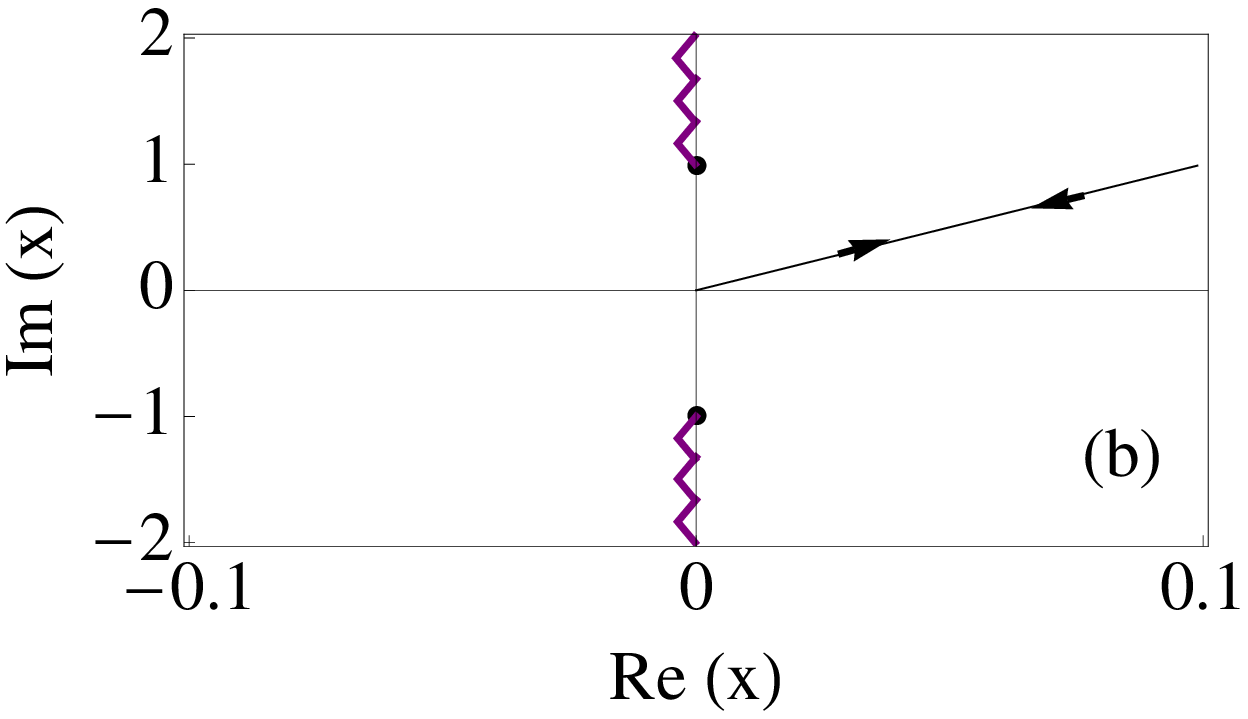}
\end{center}
\caption{\label{fig_z,x,cpr}
(Color online)
CPR process.
% with a constant detuning and a Gaussian Rabi frequency.
Branch cuts and representative trajectories in the complex (a) $z$ and (b) $x$ planes, see Eqs. (\ref{z}) and (\ref{x}), with
parameters as in Fig. \ref{fig_E}.
%$\Gamma =2\pi\times3.183$ kHz, $\Omega_{max} =2\pi\times1.592$ kHz, $a =4\times10^{8}$ s$^{-2}$,
%$\Delta_0=2\pi \times0.159$ kHz, and $t_f=1$ ms.
}
\end{figure}
%%%%%%%%%%%%%%%%%%%%%%%%%%%%%%%%%%%%%%%%%%%%%end figure%%%%%%%%%%%%%%%%%%%%%

The second type of process we consider is  CPR \cite{CPR} with constant detuning $\Delta_0>0$, and a Rabi frequency
given by a Gaussian function, 
\beqa
\label{delta_omega_cpr}
\Delta_{cpr}(t) &=& \Delta_0,
\\
\Omega_{R, cpr}(t) &=& \Omega_{max}  e^{[-a (t-t_f/2)^2]},
\eeqa
where $\Omega_{max}$ and $a$ are constants.
For this process $z$ has a constant imaginary part, $\text{Im}(z) = -4 \Gamma \Delta_{0}$.
Thus, $z(t)$ never crosses the real axis and we may choose the branch cut along (just below) the negative part of this axis,
as shown in Fig. \ref{fig_z,x,cpr} (a). Then, $|-\ra$ is the least dissipative state throughout. 
%\red{for the parameters
%$\Gamma =2\pi\times3.183$ MHz, $\Omega_{max} =2\pi\times1.592$ MHz, $a =400\times10^{12}$ s$^{-2}$,
%$\Delta=2\pi \times0.159$ MHz, and $t_f=1$ $\mu$s.}
The trajectory of $x$ moves back and forth in the first quadrant,
so the branch cuts are chosen as depicted  in Fig. \ref{fig_z,x,cpr} (b),  without adding $\pi$ to define $\alpha$.
Now $\alpha(0)\approx\alpha(t_f)\approx0$, $|+(0)\ra\approx|+(t_f)\ra\approx |{\sf e}\ra$, and   $|-(0)\ra\approx|-(t_f)\ra\approx |{\sf g}\ra$.  
The  eigenenergies behave as in Figs. \ref{fig_E} (e) and \ref{fig_E} (f).
\section{Numerical examples}
%
%
%
%
%
%
%
%%%%%%%%%%%%%%%%%%%%%%%%%%%%%%%%%%%%%%%%%%%%%begin figure%%%%%%%%%%%%%%%%%%%%%%%%%%%%%%%%%%%%%%%%%%%%%%%%%%%%%%%%%
\begin{figure}[t]
\begin{center}
\includegraphics[height=2.5cm,angle=0]{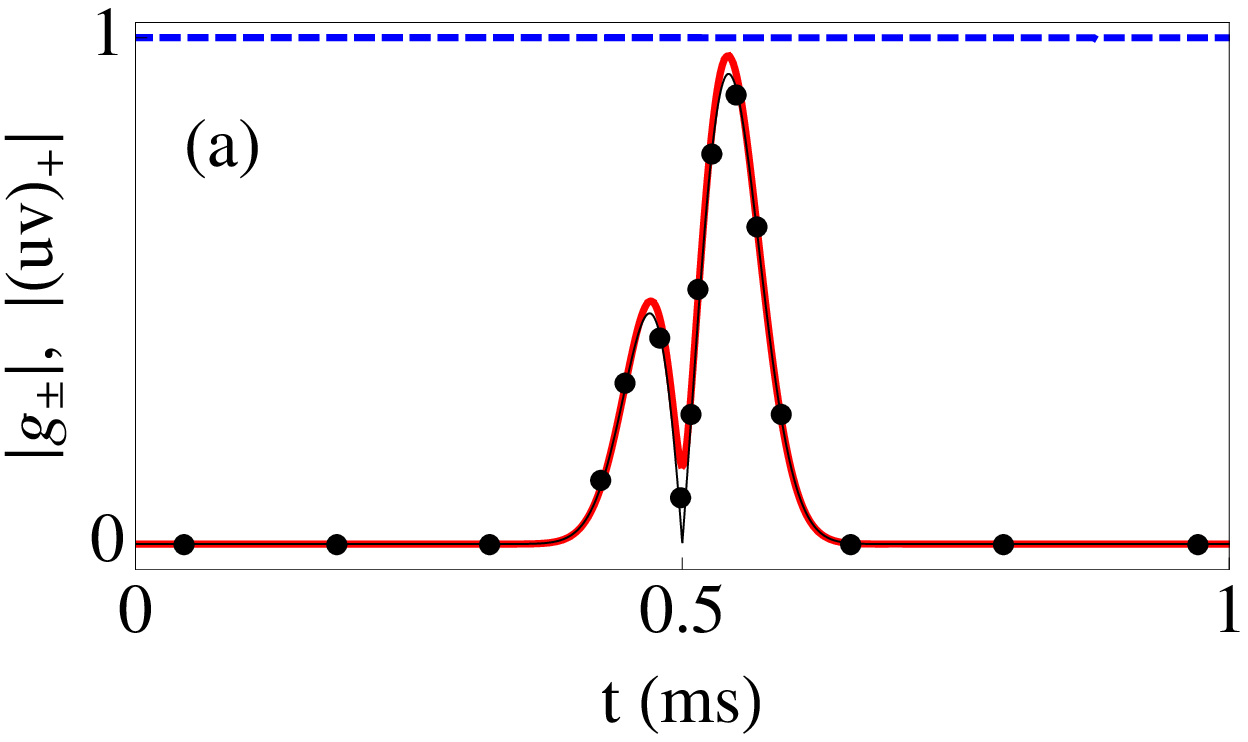}
\includegraphics[height=2.5cm,angle=0]{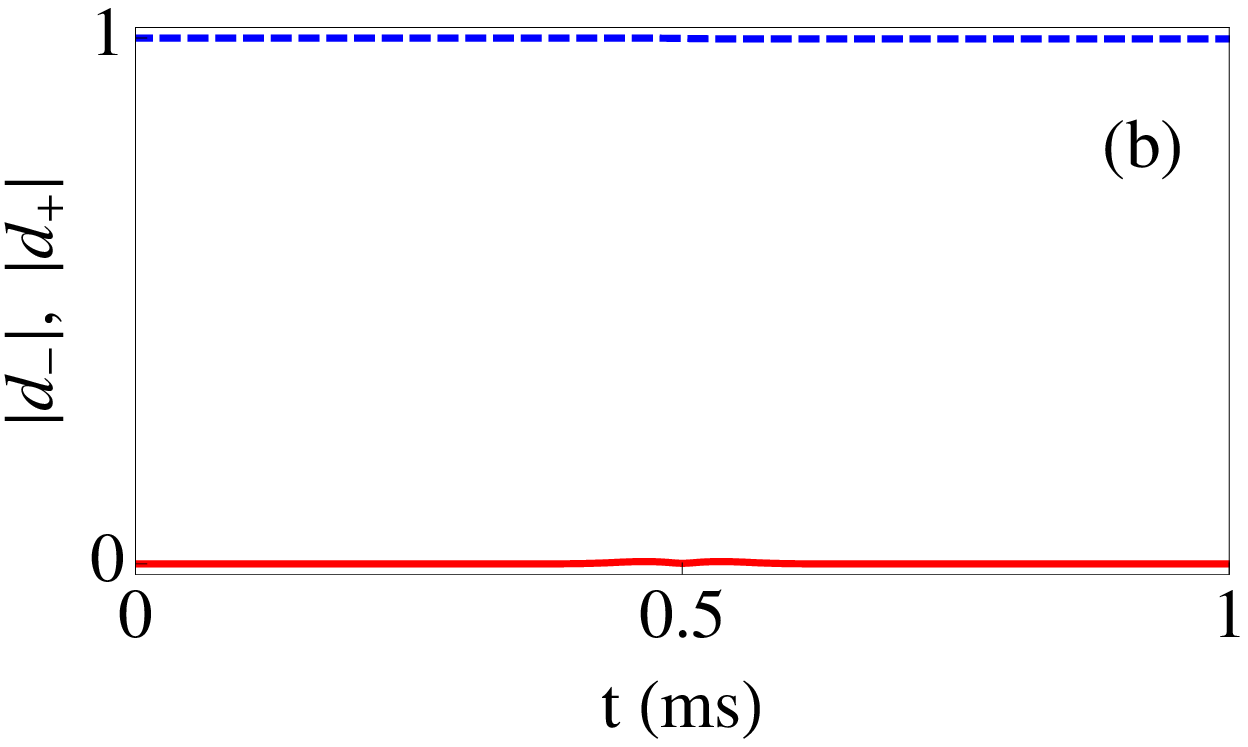}
\includegraphics[height=2.5cm,angle=0]{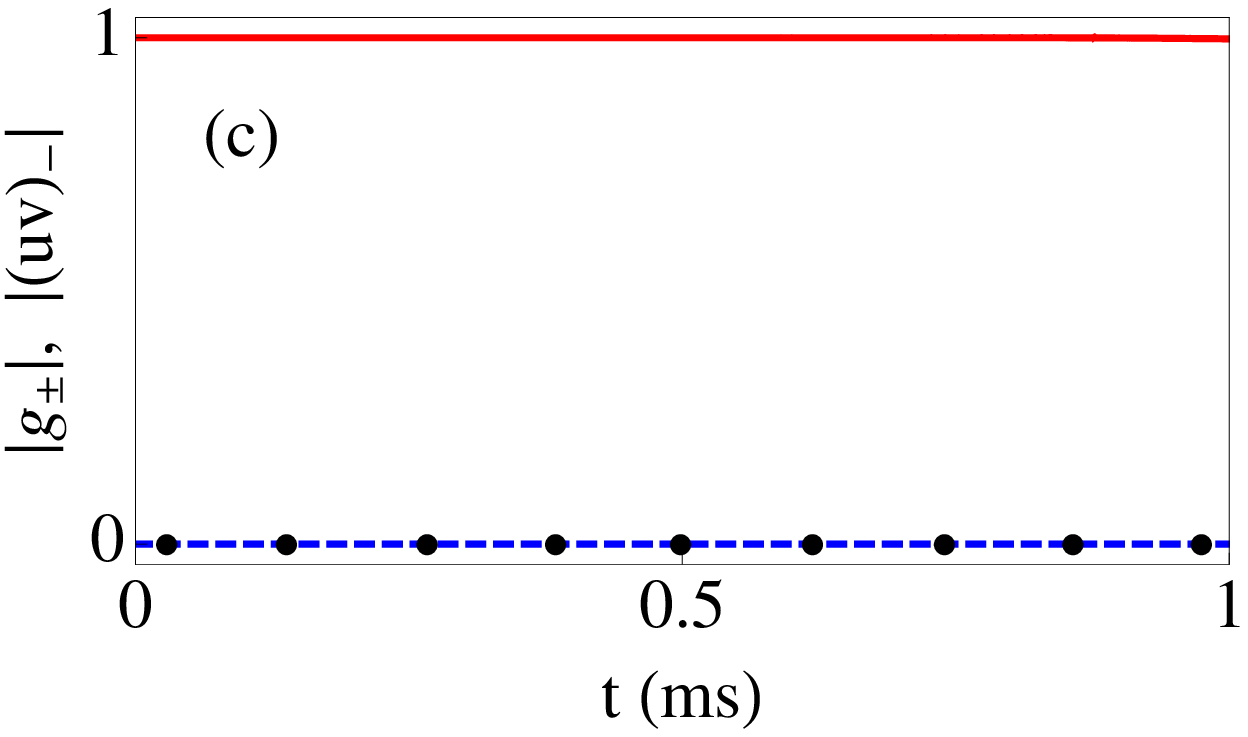}
\includegraphics[height=2.5cm,angle=0]{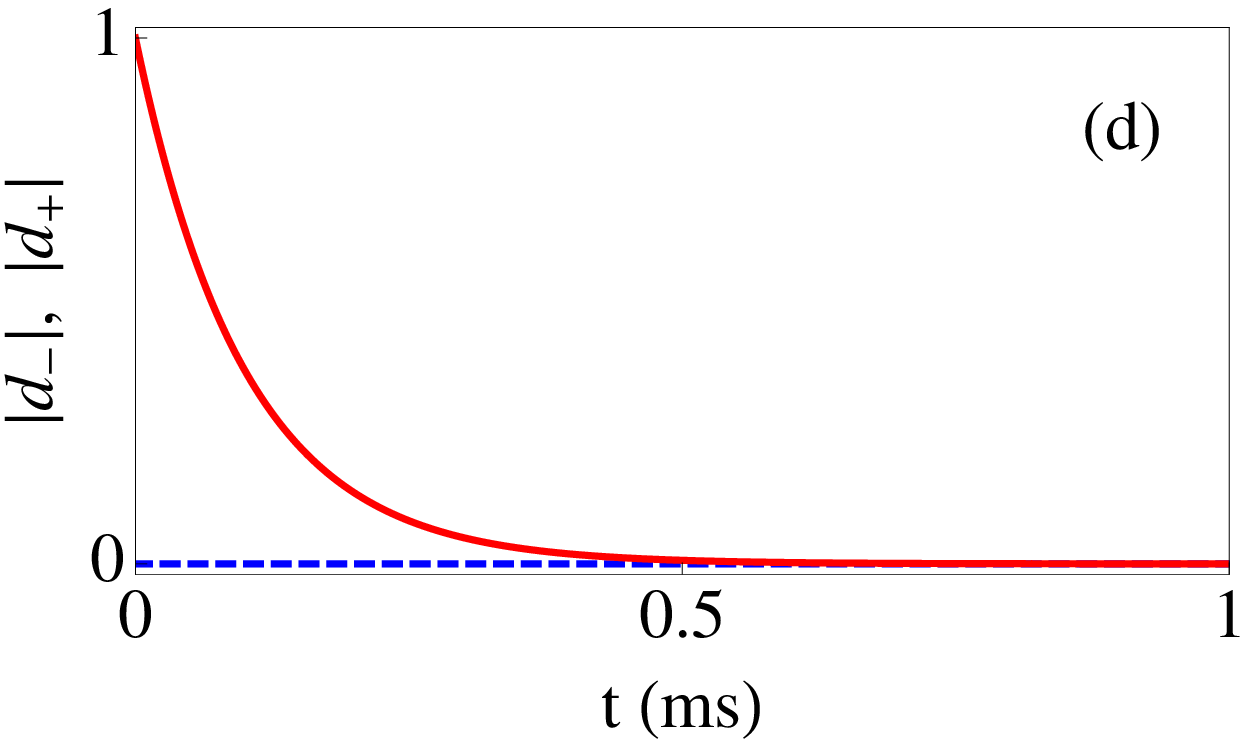}
\end{center}
\caption{\label{fig_g,d}
(Color online)
$|g_\pm|$, $|d_\pm|$ and $|(uv)_{\pm}|$ for a CPR process:
% with a constant detuning and a Gaussian Rabi frequency: 
(a,b): $|\Psi(0)\ra = |g\ra = |-(0)\ra$ (least dissipative). 
(c,d): $|\Psi(0)\ra = |e\ra = |+(0)\ra$ (most dissipative). 
The red solid line is for $|g_+|$ and $|d_+|$, and the blue dashed line for $|g_-|$ and $|d_-|$. 
The black dots 
%with or without a  black line 
are the approximation $|(uv)_\pm|$ for the state that is not
populated initially.   
%(a) $|g_+(t)|$ (red thick line) and $|g_-(t)|$ (blue thick dashed line),
%(b) $|d_+(t)|$ (red thick line) and $|d_-(t)|$ (blue thick dashed line) for $|\Psi(0)\ra = |g\ra = |-(0)\ra$,
%(c) $|g_+(t)|$ (red thick line) and $|g_-(t)|$ (blue thick dashed line), and
%(d) $|d_+(t)|$ (red thick line) and $|d_-(t)|$ (blue thick dashed line) for $|\Psi(0)\ra = |e\ra = |+(0)\ra$. 
Parameters:
$\Gamma =2\pi \times 3.183$ kHz, $\Omega_{max} =2\pi \times 3.183$ kHz, $a =4\times 10^{8}$ s$^{-2}$,
$\Delta_0 = 2\pi \times 31.831$ kHz, and $t_f=1$ ms.
}
\end{figure}
%%%%%%%%%%%%%%%%%%%%%%%%%%%%%%%%%%%%%%%%%%%%%end figure%%%%%%%%%%%%%%%%%%%%%%%%%%%%%%%%%%%%%%%%%%%%%%%%%%%%
%
%%%%%%%%%%%%%%%%%%%%%%%%%%%%%%%%%%%%%%%%%%%%%begin figure%%%%%%%%%%%%%%%%%%%%%%%%%%%%%%%%%%%%%%%%%%%%%%%%%%%%%%%%%
\begin{figure}[h]
\begin{center}
\includegraphics[height=4.0cm,angle=0]{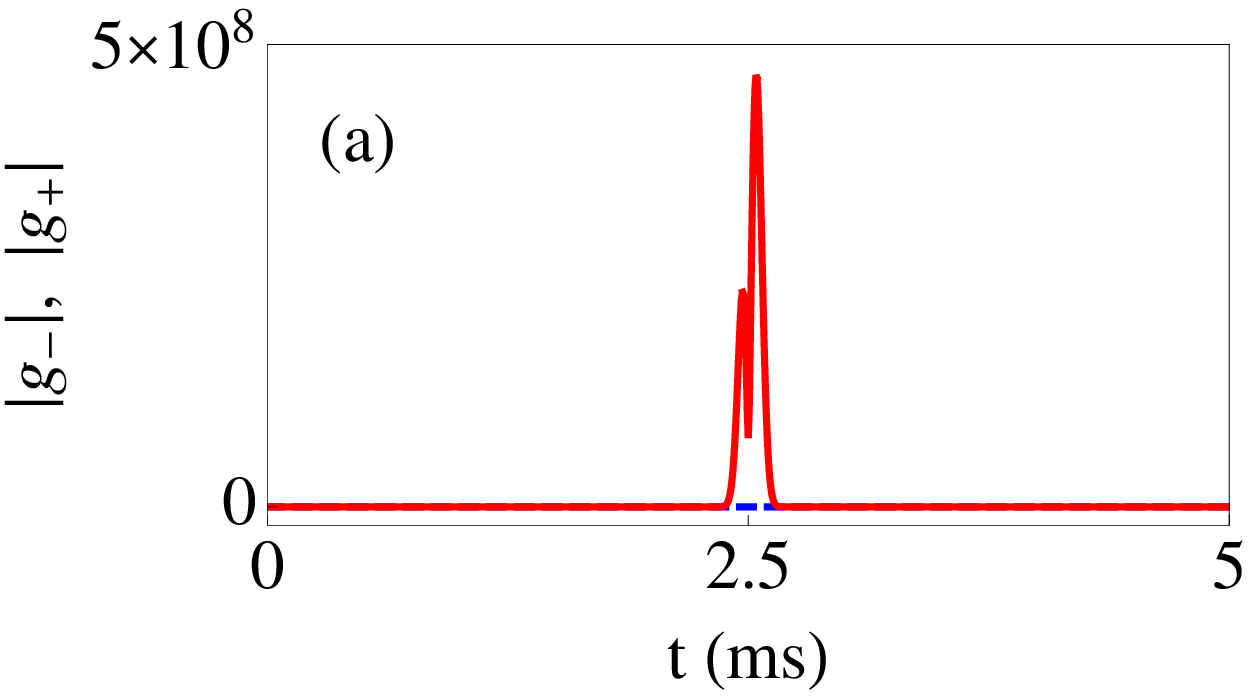}
\includegraphics[height=3.8cm,angle=0]{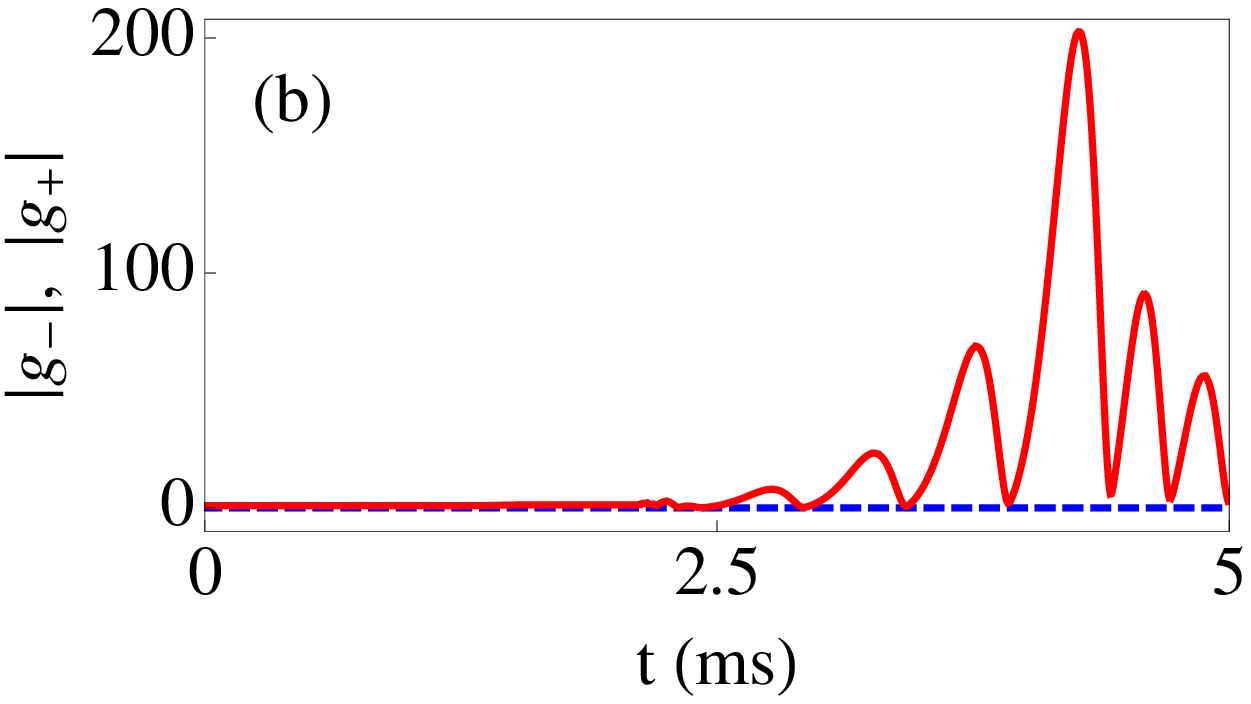}
\end{center}
\caption{\label{fig_gfail}
(Color online)
$|g_+(t)|$ (red solid line) and $|g_-(t)|$ (blue dashed line) for a CPR process
%with a constant detuning and a Rabi frequency given by a Gaussian function
when
(a) $|\Psi(0)\ra = |{\sf g}\ra = |-(0)\ra$ and (b) $|\Psi(0)\ra = |{\sf e}\ra = |+(0)\ra$,
for the parameters:
$\Gamma =2\pi \times 3.183$ kHz, $\Omega_{max} =2\pi \times 3.183$ kHz, $a =4\times 10^{8}$ s$^{-2}$,
$\Delta_0 = 2\pi \times 31.831$ kHz, and $t_f=5$ ms.
}
\end{figure}
%%%%%%%%%%%%%%%%%%%%%%%%%%%%%%%%%%%%%%%%%%%%%end figure%%%%%%%%%%%%%%%%%%%%%%%%%%%%%%%%%%%%%%%%%%%%%%%%%%%%
%
%%%%%%%%%%%%%%%%%%%%%%%%%%%%%%%%%%%%%%%%
\begin{figure}[t]
\begin{center}
\includegraphics[height=2.32cm,angle=0]{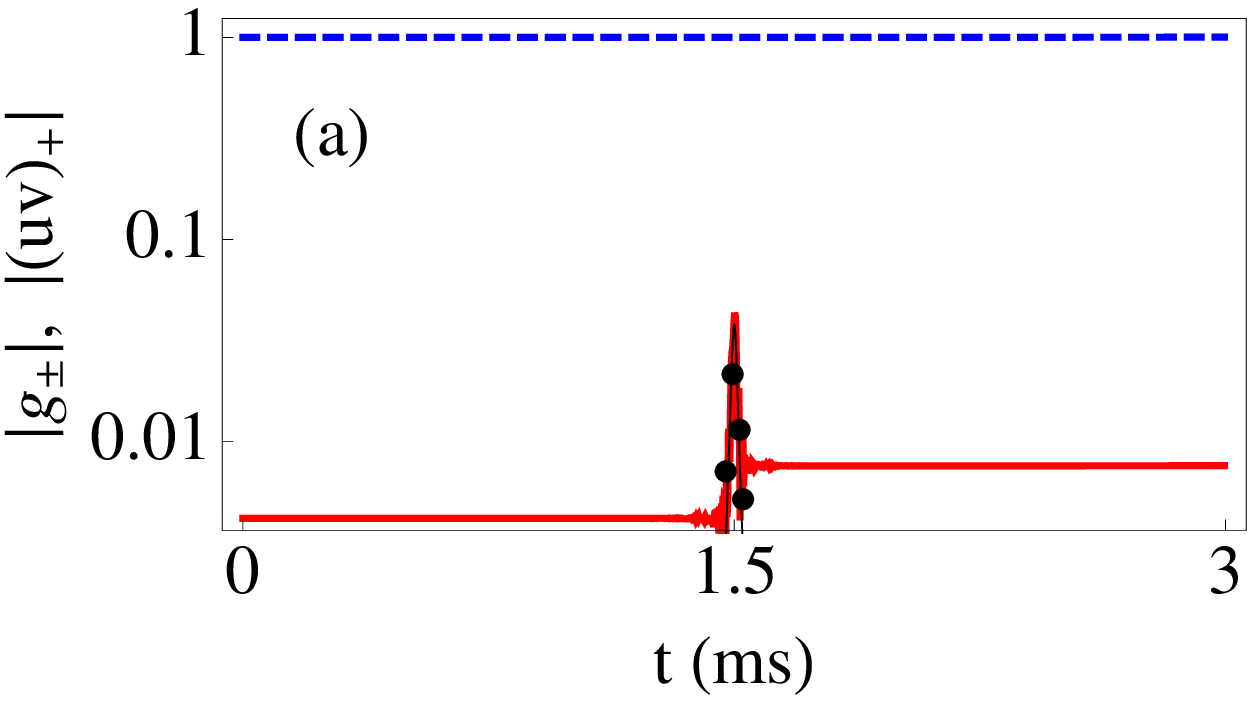}
\includegraphics[height=2.32cm,angle=0]{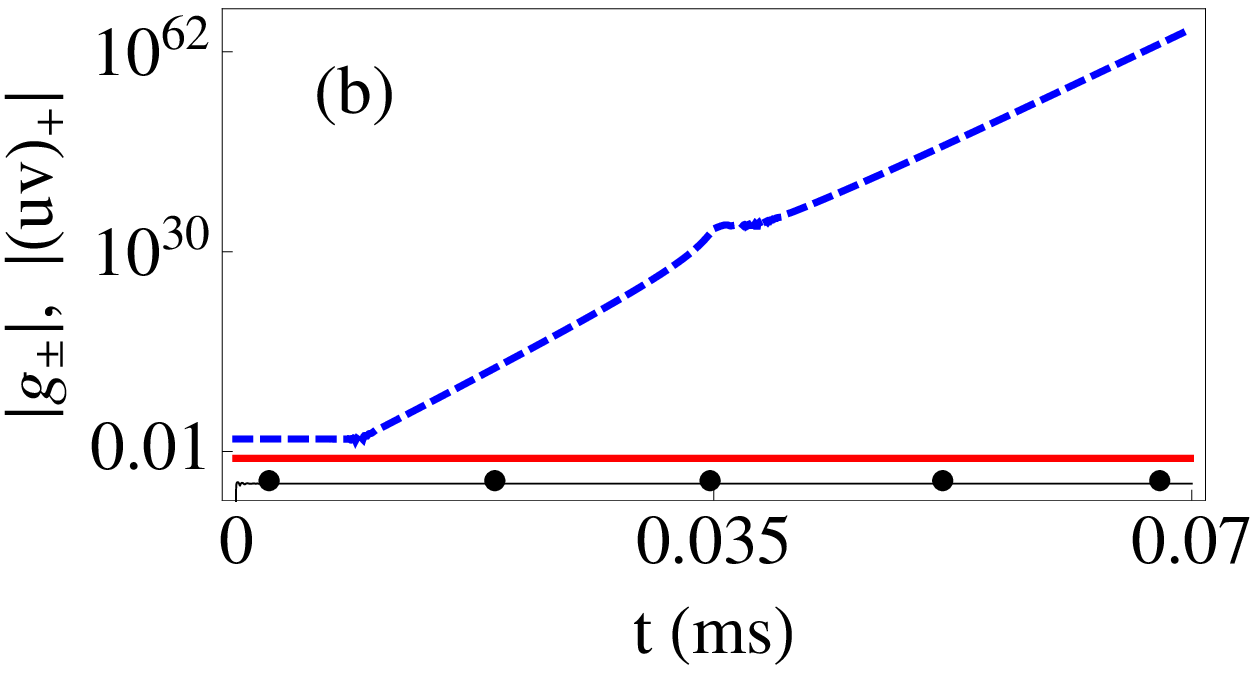}
\includegraphics[height=2.32cm,angle=0]{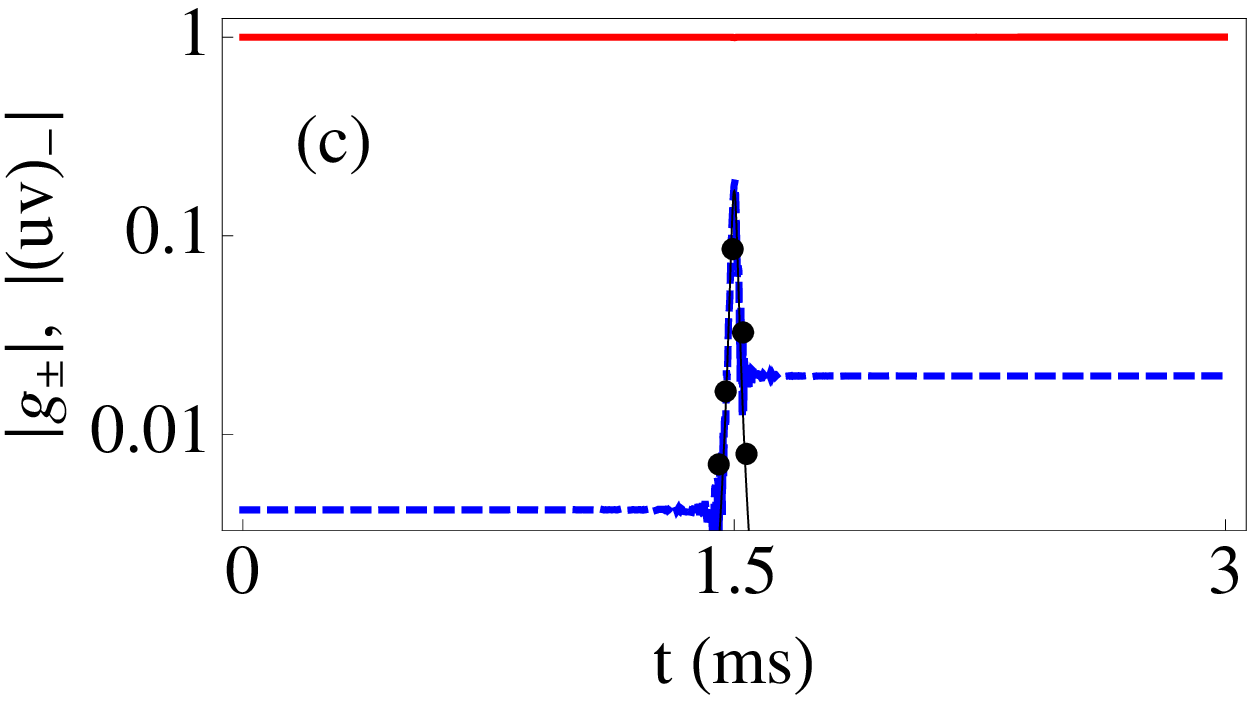}
\includegraphics[height=2.32cm,angle=0]{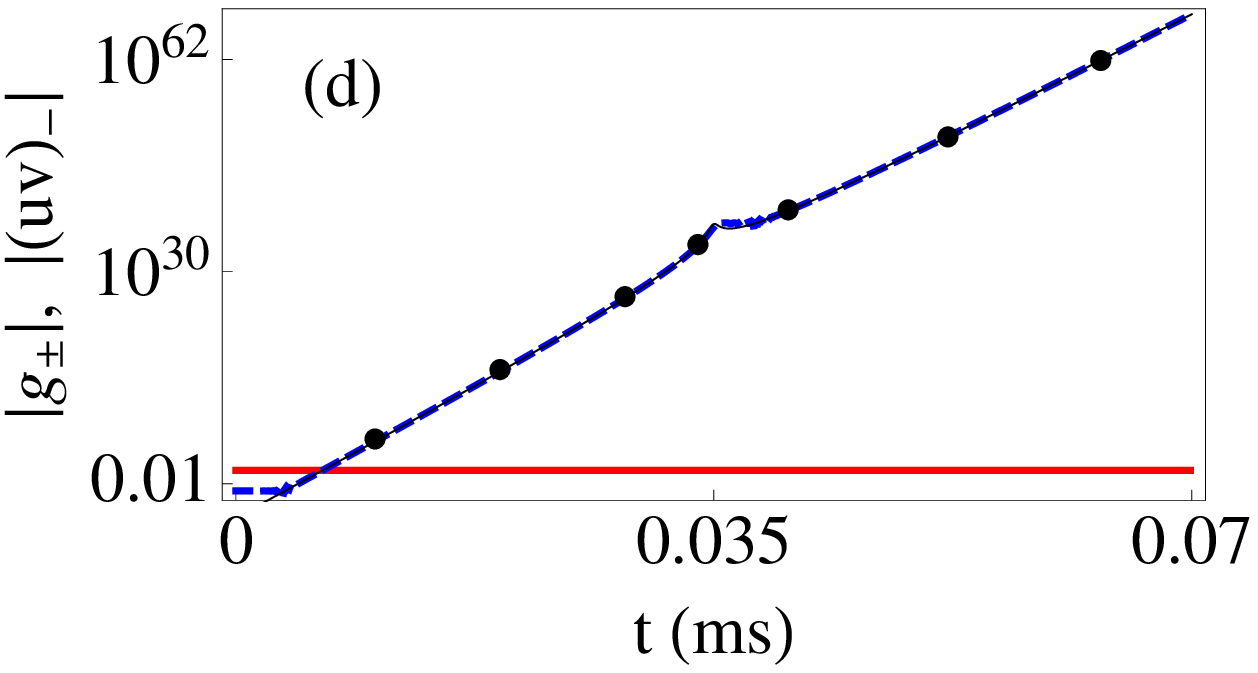}
\end{center}
\caption{\label{fig_peqgran}
(Color online)
$|(uv)_\pm(t)|$ and  $|g_\pm(t)|$ for two  Landau-Zener processes. 
$|g_+|$:  red solid line;  $|g_-|$: blue dashed line;    
$|(uv)_\pm|$: black line with dots  (for the state which is initially unoccupied).   
Initial states: 
$|\Psi(0)\ra =|{\sf e}\ra = -|-(0)\ra$ in (a,b), and $|\Psi(0)\ra =|{\sf g}\ra=|+(0)\ra$ in (c,d). 
(a,c): 
$\Gamma < 2 \Omega_0$
with 
$\Gamma =2\pi \times 0.159$ kHz, $\Omega_0 =2\pi \times 79.578$ kHz, $b =4 \times 10^{10}$ s$^{-2}$,
and $t_f=3$ ms.
(b,d):
$\Gamma > 2 \Omega_0$
with  
$\Gamma =2\pi \times 799.775$ kHz, $\Omega_0 =2\pi \times 79.578$ kHz, $b =9 \times 10^{12}$ s$^{-2}$,
and $t_f=0.07$ ms. 
}
\end{figure}
%%%%%%%%%%%%%%%%%%%%%%%%%%%%%%%%%%%%%%%%%%%%%end figure%%%%%%%%%%%%%%%%%%%%%%%%%%%%%%%%%%%%%%%%%%%%%%%%%%%%   
%%%%%%%%%%%%%%%%%%%%%%%%%%%%%%%%%%%%%%%%%%%%%%begin figure%%%%%%%%%%%%%%%%%%%%%%%%%%%%%%%%%%%%%%%%%%%%%%%%%%%%%%%%%
\begin{figure}[h]
\begin{center}
\includegraphics[height=4.0cm,angle=0]{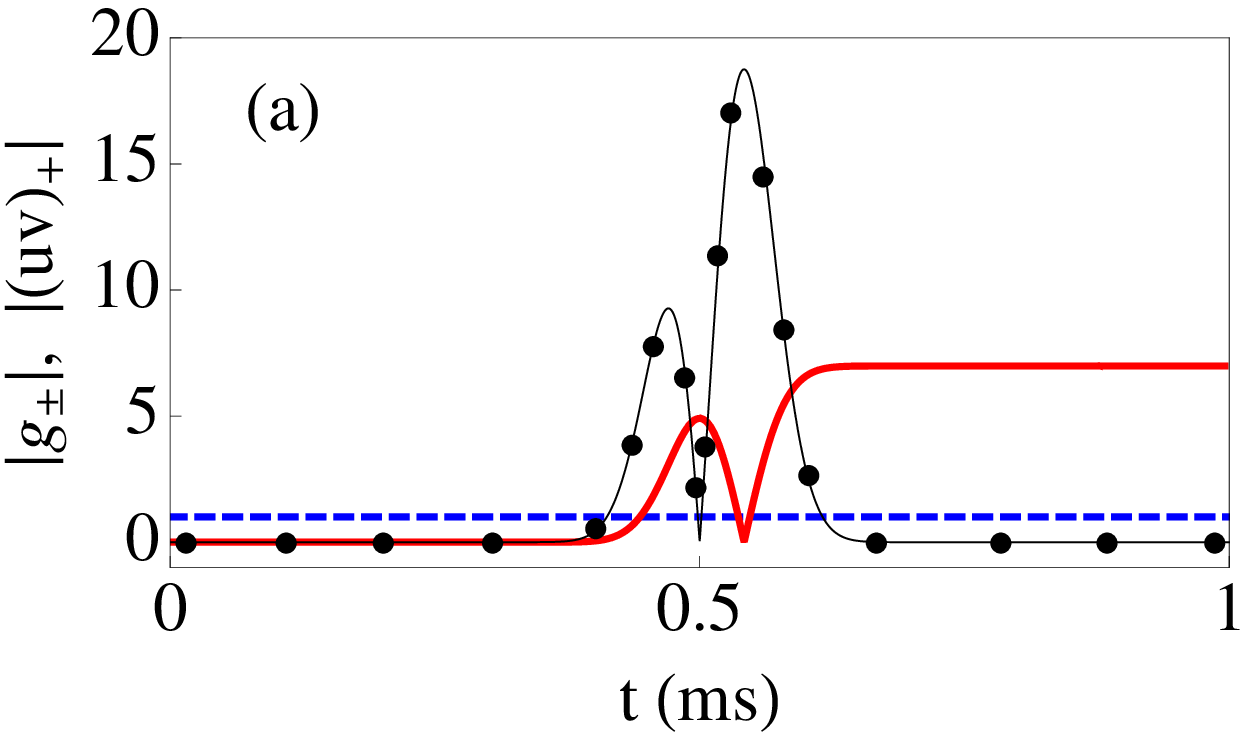}
\includegraphics[height=4.0cm,angle=0]{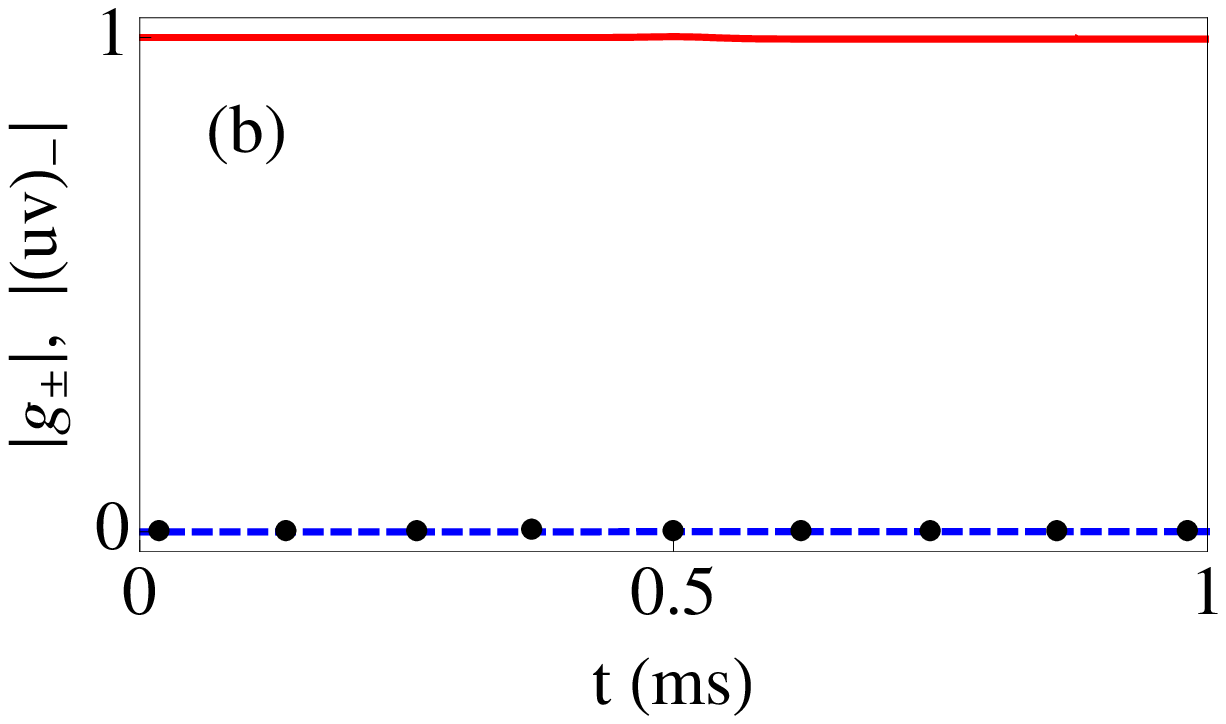}
\end{center}
\caption{\label{CPRbad}
(Color online)
$|g_+(t)|$ (red solid line), $|g_-(t)|$ (blue dashed line) and $|(uv)_{\pm}|$ (black dots with or without a  black line) for a CPR process
%ning and a Rabi frequency given by a Gaussian function
when
(a) $|\Psi(0)\ra = |\sf{g}\ra = |-(0)\ra$ (least dissipative) and (b) $|\Psi(0)\ra = |{\sf e}\ra = |+(0)\ra$ (most dissipative),
for the parameters:
$\Gamma =2\pi \times 3.183$ kHz, $\Omega_{max} =2\pi \times 0.159$ kHz, $a =4\times 10^{8}$ s$^{-2}$,
$\Delta_0 = 2\pi \times 2$ Hz, and $t_f=1$ ms.
}
\end{figure}
%%%%%%%%%%%%%%%%%%%%%%%%%%%%%%%%%%%%%%%%%%%%%end figure%%%%%%%%%%%%%%%%%%%%%%%%%%%%%%%%%%%%%%%%%%%%%%%%%%%%

In this section we shall provide, based on the two-level model, some examples to illustrate different features of adiabaticity
for NH systems,  defined in terms of the amplitudes $g_{n}(t)$. We shall mostly pay attention to properties that differ from 
the ones of Hermitian systems. 
 
Figure \ref{fig_g,d} compares, for a CPR process,  the rather different behavior of $|g_\pm|$ (left panels) and $|d_\pm|$ (right panels). In Fig. \ref{fig_g,d} (a,b) the initial state is the ground state which evolves adiabatically as the least dissipative state. 
Fig. \ref{fig_g,d} (a) for $|g_\pm|$ shows an interesting feature of NH systems, namely, that one state may remain adiabatic, 
whereas the other one does not. This is not reflected as clearly in 
Fig. \ref{fig_g,d} (b) for $|d_\pm|$.
Figs. \ref{fig_g,d} (c,d) correspond to the atom starting in the most dissipative state. Fig. \ref{fig_g,d} (c) for $|g_\pm|$ shows that 
for the time considered both states remain perfectly adiabatic. However the $|d_+|$ coefficient decays strongly 
because of spontaneous decay, see Fig. \ref{fig_g,d} (d), so a ratio $|d_-|/|d_+|$ is not  a faithful indicator of adiabaticity.  Nevertheless, 
as pointed out earlier, these coefficients are actually quite relevant, in particular at $t_f$, because
here the states $|\pm(t_f)\ra$ become 
orthonormalized and coincide with the bare basis of excited and ground atomic states. 
%In this regard the 
%strong exponential suppression of the
%final probability  of the most dissipative state $|+\ra$  ($\approx|e\ra$ at $t_f$)    
%could make its non-adiabatic excitations irrelevant at least with regard to the final state. 
%

On a different thread, note that in the examples of Figs. \ref{fig_g,d} (a) and \ref{fig_g,d} (c),  $|(uv)_\pm|$, see Eqs. (\ref{uv}-\ref{uv_n}), 
are very good approximations to $|g_\pm|$ for the initially unoccupied states
(the subscript in  $|(uv)_\pm|$ specifies which amplitude, $|g_\pm|$, is approximated).

% ... provides an exale of another peculiar phenomenon: the fact that longer process times may actually spoil
%adiacity for NH systems. In the example shown, for a CPR protocol,  the system starts in the most-dissipative state.
%This effect is not seen when the system starts (and stays) in the least-dissipative state. 

Fig. \ref{fig_gfail} is about a CPR process with a final time five times larger than in the previous figure. 
Contrary to Hermitian systems, longer process times may actually spoil
adiabaticity for NH systems.  
Fig. \ref{fig_gfail} (b) shows,   compare to Fig. 4 (c), that when the system  starts in the most dissipative state, $|+\ra$, 
it does not remain adiabatic if the time is large enough.    
Contrast this also to Fig. \ref{fig_gfail} (a), where  the system starts and stays adiabatic
in the least dissipative state, $|-\ra$, while $|+\ra$ is excited.  
%

%%%%%%%%%%%%%%%%%%%%%%%%%%%%%%%%%%%%%%%%%%%%%begin figure%%%%%%%%%%%%%%%%%

The approximations $|(uv)_\pm|$ in Eq. (\ref{uv_n}) are depicted in Fig. \ref{fig_peqgran} in logarithmic scale for Landau-Zener  processes with
decay 
for $\Gamma<2\Omega_0$ (left panels) and $\Gamma>2\Omega_0$ (right panels).   
In general the criterion $|(uv)_\pm|<<1$ 
avoids the gross pitfalls of simpler choices at crossings of the real and imaginary parts of the energies, 
see Eqs. (\ref{uv'}) and (\ref{uv''}), as long as a fully degenerate point (when both real and imaginary parts are equal) is not crossed.  
%Considering its simplicity, it provides at least a rough guidance for the breakdown of adiabaticity
%far from a degenerate point in parameter space. 
%

In general though, $-(uv)_\pm(t)$ do not reproduce  $g_\pm(t)$ accurately, even when the condition of  first order perturbation
theory,
$g_m(t)\approx 1$, holds. A clear example taken from CPR is depicted  in Fig. \ref{CPRbad} (a), where the  
remainder integral, $\int_0^t v_+ du_+ $, see Eq. (\ref{parts}),  is not 
small, so $-(uv)_+(t)$ is quite different from $g_+(t)$ even though $g_-(t)\approx1$ during the process. 
Contrast the failure of $-(uv)_+$ in Fig. \ref{CPRbad} (a) with the accurate fitting in Fig. \ref{fig_g,d} (a).   
Integration by parts 
provides a formal series in powers of $\omega_{nm}(t)$, as shown in  Appendix A, where the only critical points are the end points,
but other points may play an important role. 
The  approximation $g_n(t)\approx -(uv)_n(t)$, from the first term in Eq. (\ref{parts}), can also 
be found by assuming  $u_n(t'<t)\approx u_n(t)$ in $g_n(t)=-\int_0^t u_n dv_n$
(we assume also that the contribution at $t=0$ is negligible). 
This substitution though is not always permissible. Take for example $t=t_f$ in Figs. \ref{fig_g,d}, and \ref{CPRbad}. Fig.    
\ref{CPRcomp} demonstrates that the oscillation or otherwise of $e^{iW_{+-}}$ makes the $u_+$ contribution around 
$t_f/2$ either irrelevant (in Fig. \ref{fig_g,d}) or quite significant (in Fig. \ref{CPRbad}). In the later case, the approximation based only on the critical point at $t_f$ cannot be accurate.    
An alternative view making use of the complex-time plane to perform the integrals is provided in Appendix B. 
In general, accurate approximations of the $g_\pm(t)$ requires contour deformations 
in the complex time plane \cite{Guerin2002,DG} to identify and take into account contributions from all relevant eigenvalue degeneracies and other critical points. As well, crossings of Stokes lines \cite{Berry89}  determine changes in the asymptotic behavior of the amplitudes \cite{BU}. While this type of analysis is possible for simple specific models  and protocols \cite{DG,Berry89,BU}, it may easily become  
intractable for moderately complex systems (such as a generic three level system \cite{DG}) due to the proliferation of singularities \cite{DG}.  An open question then is to bridge the gap between
a simple condition like (\ref{uv_n}) and more accurate conditions  in generic cases.  
% 
%%%%%%%%%%%%%%%%%%%%%%%%%%%%%%%%%%
\begin{figure}[t]
\begin{center}
\includegraphics[height=2.1cm,angle=0]{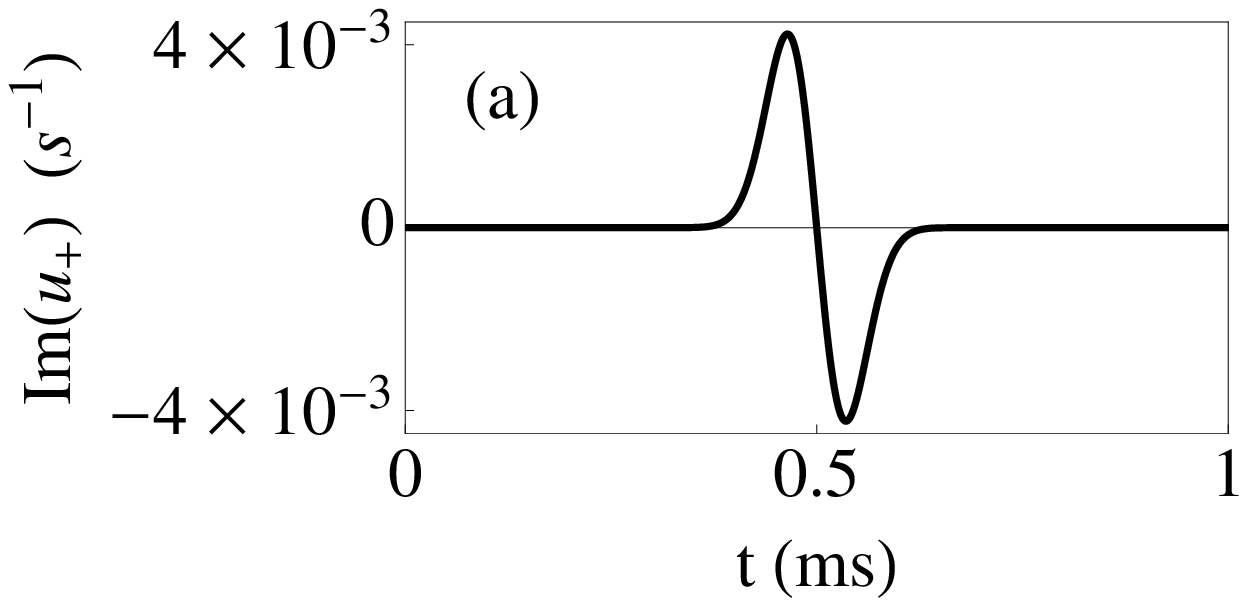}
\includegraphics[height=2.1cm,angle=0]{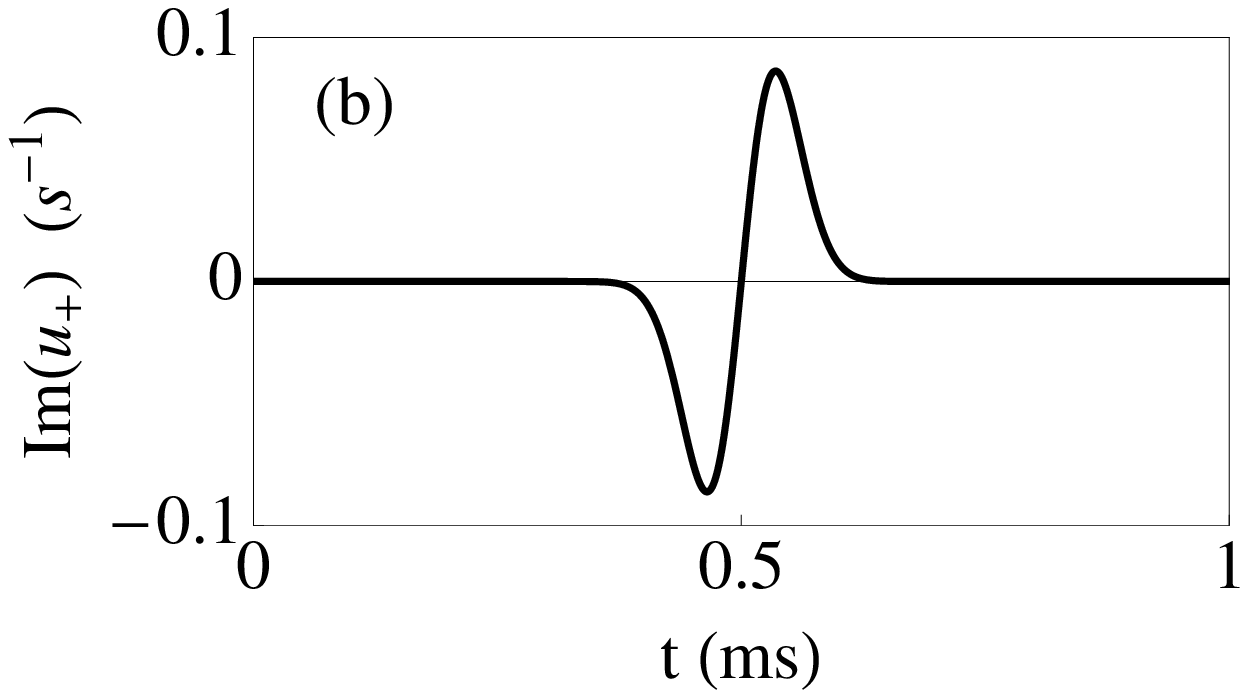}
\includegraphics[height=2.1cm,angle=0]{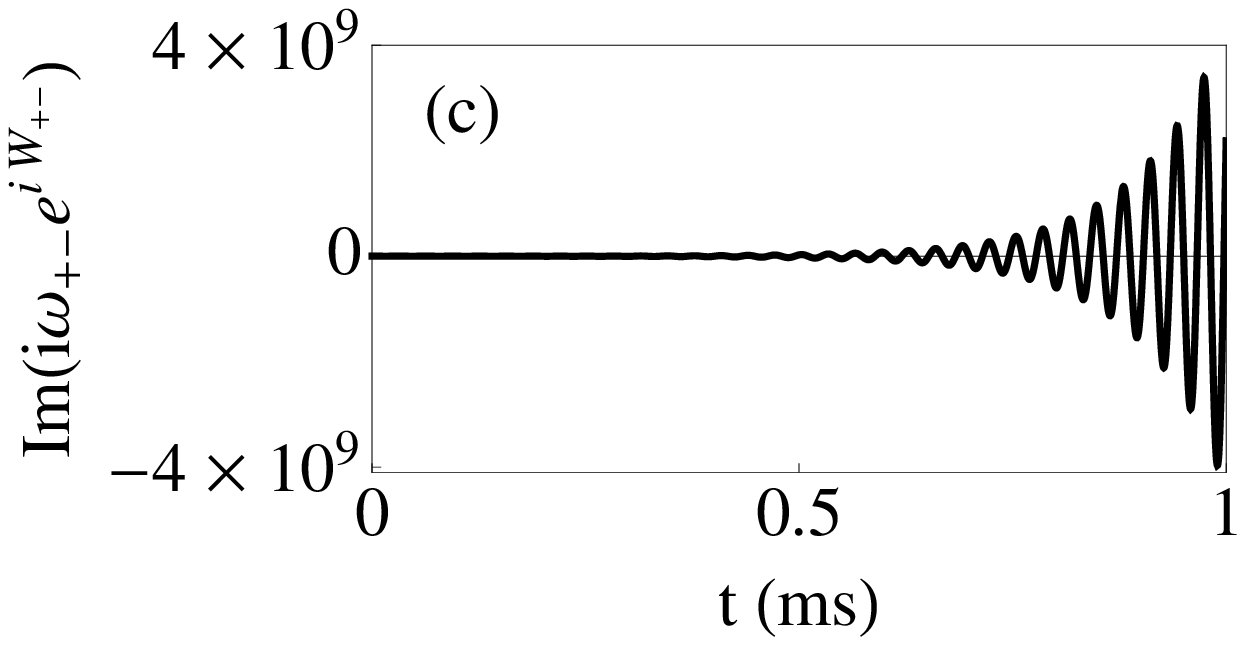}
\includegraphics[height=2.1cm,angle=0]{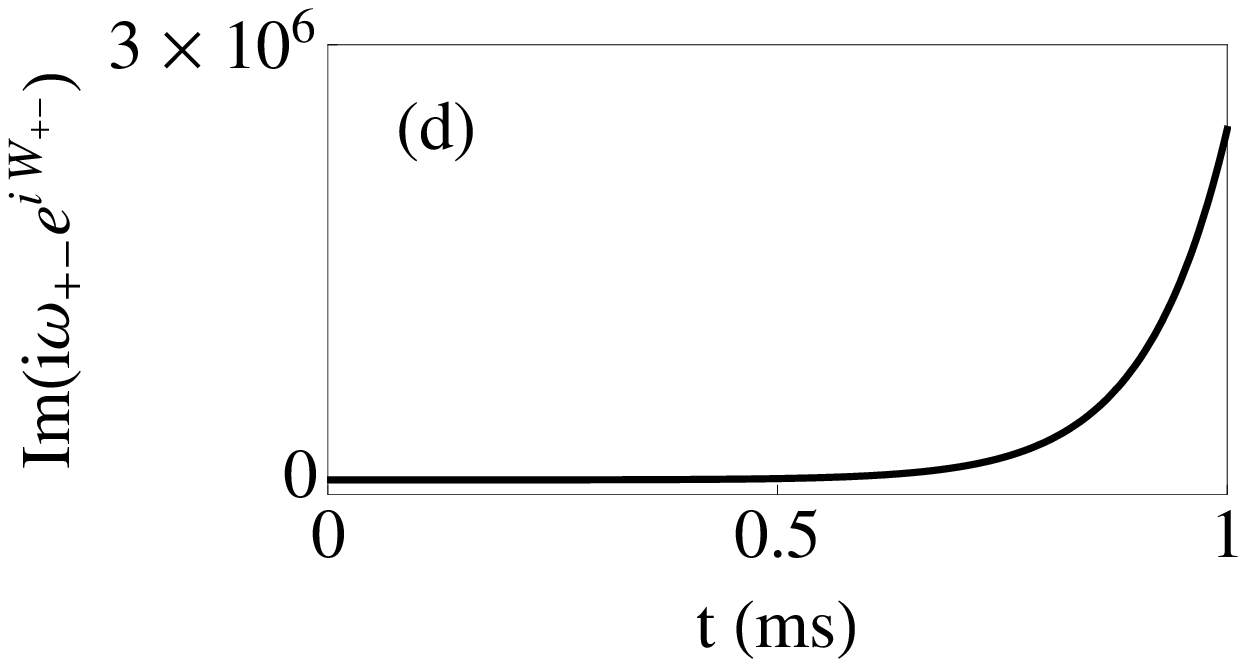}
\includegraphics[height=2.1cm,angle=0]{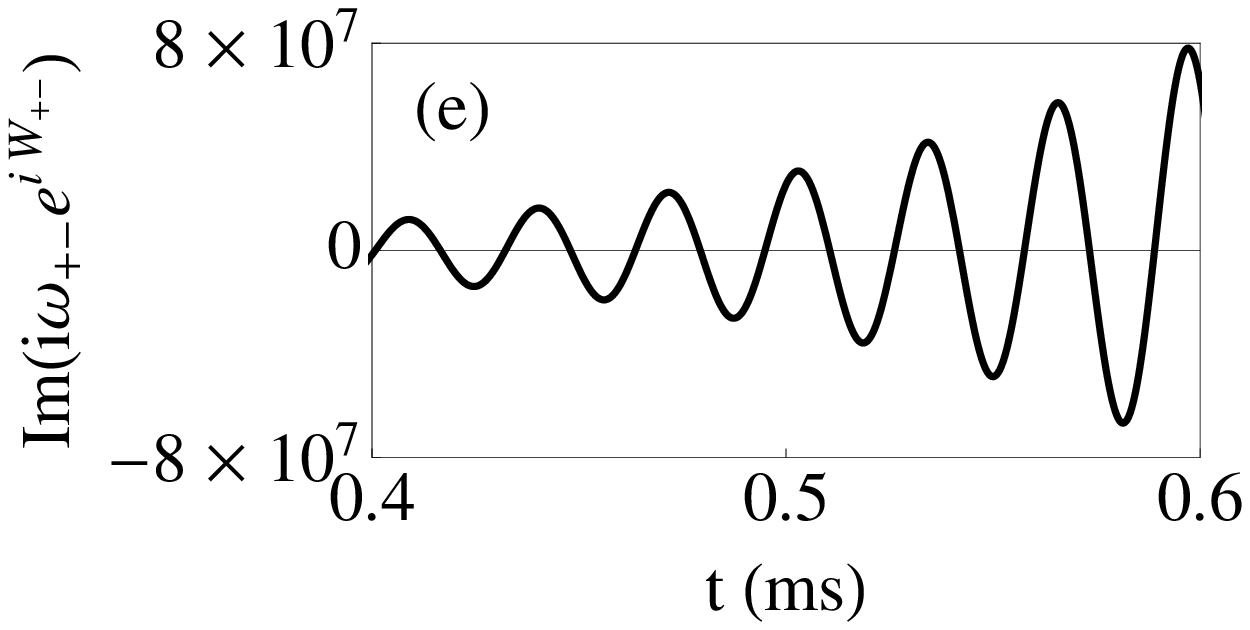}
\includegraphics[height=2.1cm,angle=0]{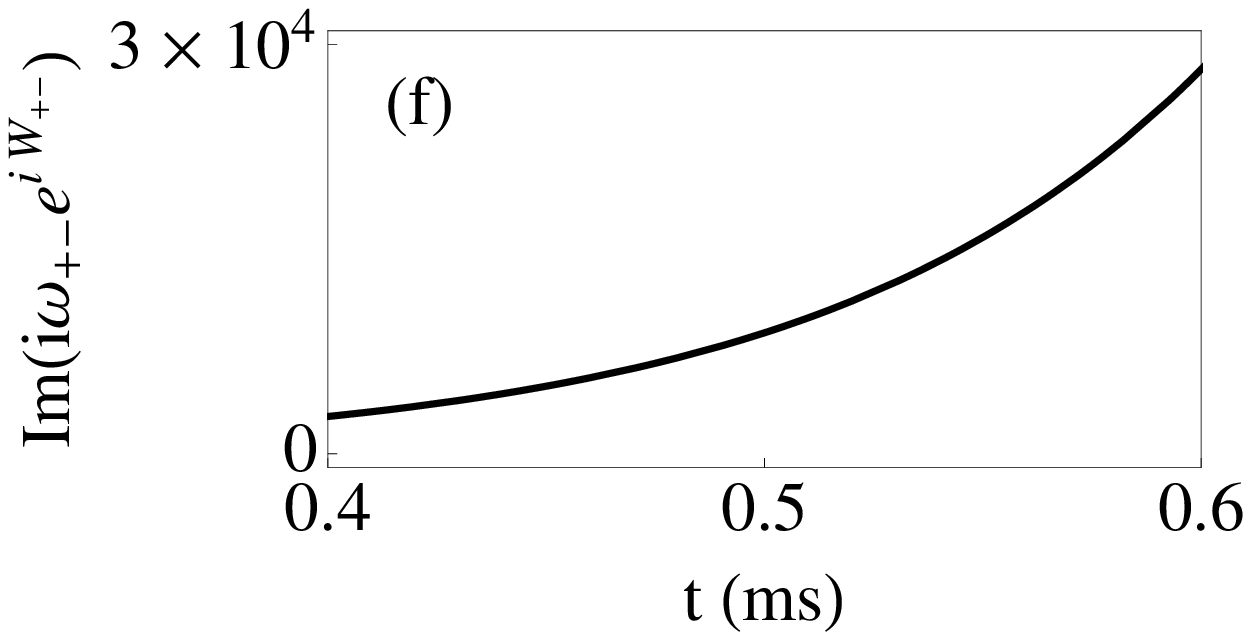}
\end{center}
\caption{\label{CPRcomp}
Comparison of imaginary parts of integrand terms of $g_+(t)$, see Eqs. (\ref{aco}) and (\ref{uv}), for the CPR processes of Fig. 4 (a) (left panels) 
and Fig. \ref{CPRbad} (a) (right panels). (a,b):  Im($u_+$); (c,d): Im$(i\omega_{+-}e^{iW_{+-}})$; (e,f): zoom of (c,d) around the central time. 
The corresponding figures for the real parts are qualitatively quite similar. 
}
\end{figure}
%%%%%%%%%%%%%%%%%%%%%%%%%%%%%%%%%%%%%%%%%%%%%end figure%%%%%%%%%%%%%%%%%%%%%%%%%
%
%
%
%
%
%
\section{Discussion}
Adiabaticity is a key concept in quantum physics and its generalization  to systems described by non-Hermitian
dynamics 
requires the analysis of several possible formal extensions of the populations conserved 
for adiabatic dynamics in Hermitian systems. 
We have  
singled out among them the one that best identifies adiabatic following because it  remains adiabatically invariant. 
Examples to illustrate its behavior have been drawn from 
CPR and LZ processes.     
A simple approximate expression has been also worked out by perturbation theory and partial integration, as well as 
higher orders in inverse
powers of the transition frequency, for studying the adiabaticity of states 
different from the one initially occupied. It appears as a natural generalization of the usual condition for Hermitian systems. 
Its Hermitian counterpart is not infallible 
\cite{MS,Duetal,PRLS,Ortigoso}, so an accurate  performance cannot be expected in general, as shown in the examples.  
This suggests many directions for future work: A systematic analysis and prediction of its possible failures is needed. 
In  addition to the 
reasons found in its  Hermitian counterpart, other  elements have to be considered, such as the occurrence of NH degeneracies \cite{BU,UMM}.   
The simple approach to NH adiabaticity followed here, in exact or approximate forms, should also be contrasted with
alternative views both conceptually and for specific applications. For example, in CPR, adiabaticity has been discussed 
in terms of the eigenstates of the Hermitian Hamiltonian without decay (with $\Gamma=0$ in Eq. (\ref{hami})) instead       
of the full Hamiltonian \cite{CPR}. 
%Investigating the relation of this approach with the  more fundamental method followed here
%will be worthwhile.    

Further applications or extensions of  this work may be in fields such as dissipative Master equations
\cite{LGJ,AT},  superadiabatic treatments \cite{Joye,guerin,ICM},
time-dependent dissipation rates \cite{DG}, or 
non-Hermitian quantum adiabatic computation \cite{NB}.
The formalism and concepts are also applicable beyond quantum physics, for example to treat coupled waveguides \cite{TDL}. 
\section*{Acknowledgements}
We are grateful to M. V. Berry for commenting on the original manuscript
and to A. Peralta Conde and D. Sokolovski  for discussions.  
We acknowledge funding by Grants No. IT472-10, 
FIS2009-12773-C02-01, and  
the UPV/EHU Program
UFI 11/55. S.I. acknowledges support from the Basque Government
(Grant No. BFI09.39).
\appendix
%
%
%
%
%
%
%
%%%%%%% Appendix
%
%
%
%
\section{Second and third order approximations of the $g_n(t)$}
Integrating the second term in Eq. (\ref{parts}) again by parts, 
as $\int_0^t v_n du_n = \int_0^t u_{1,n} dv_{1,n}$, where
\beqa
u_{1,n}&=&\frac{\la \dot{\widehat{n}}(t')|\dot{m}(t')\ra + \la \widehat{n}(t')|\ddot{m}(t')\ra }{[i\omega_{nm}(t')]^2}
\nonumber \\
&-&  \frac{ \la \widehat{n}(t')|\dot{m}(t')\ra  i \dot{\omega}_{nm}(t')}{[i\omega_{nm}(t')]^3}
\nonumber \\
dv_{1,n}&=&i\omega_{nm}(t')e^{iW_{nm}(t')} dt',
\nonumber
\eeqa  
we get
\beqa
%\label{parts2}
g_n(t)  &=&  \Bigg\{ -\frac{\la \widehat{n}(t')|\dot{m}(t')\ra}{i\omega_{nm}(t')}
\nonumber\\
&+&    \frac{\la \dot{\widehat{n}}(t')|\dot{m}(t')\ra + \la \widehat{n}(t')|\ddot{m}(t')\ra }{[i\omega_{nm}(t')]^2}
\nonumber \\
&-&  \frac{ \la \widehat{n}(t')|\dot{m}(t')\ra  i \dot{\omega}_{nm}(t')}{[i\omega_{nm}(t')]^3} \Bigg\} e^{iW_{nm}(t')}  \Bigg|_0^t
\nonumber\\
&-&\int_0^t v_{1,n} du_{1,n}.
\nonumber
\eeqa
We may integrate by parts the remainder integrals that appear at each step. 
First we rewrite $\int_0^t v_{1,n} du_{1,n} =  \int_0^t u_{2,n} dv_{2,n}$, with 
\beqa
%\label{uv}
u_{2,n}&=&\frac{\la \ddot{\widehat{n}}(t')|\dot{m}(t')\ra + 2 \la \dot{\widehat{n}}(t')|\ddot{m}(t')\ra
+\la \widehat{n}(t')|\dddot{m}(t')\ra}{[i\omega_{nm}(t')]^3}
\nonumber \\
&-&  \frac{[\la \dot{\widehat{n}}(t')|\dot{m}(t')\ra + \la \widehat{n}(t')|\ddot{m}(t')\ra] 3i \dot{\omega}_{nm}(t')}
{[i\omega_{nm}(t')]^4}
\nonumber \\
&-& \frac{\la \widehat{n}(t')|\dot{m}(t')\ra i  \ddot{\omega}_{nm}(t')}{[i\omega_{nm}(t')]^4}
- \frac{\la \widehat{n}(t')|\dot{m}(t')\ra 3 \dot{\omega}_{nm}^2(t')}{[i\omega_{nm}(t')]^5}
\nonumber \\
dv_{2,n}&=&i\omega_{nm}(t')e^{iW_{nm}(t')} dt'.
\nonumber
\eeqa  
Thus,
\beqa
%\label{parts3}
g_n(t)  &=&  \Bigg\{ -\frac{\la \widehat{n}(t')|\dot{m}(t')\ra}{i\omega_{nm}(t')}
\nonumber\\
&+&\frac{\la \dot{\widehat{n}}(t')|\dot{m}(t')\ra + \la \widehat{n}(t')|\ddot{m}(t')\ra }{[i\omega_{nm}(t')]^2}
\nonumber \\
&-&\frac{\la \widehat{n}(t')|\dot{m}(t')\ra  i \dot{\omega}_{nm}(t')
-\la \ddot{\widehat{n}}(t')|\dot{m}(t')\ra}{[i\omega_{nm}(t')]^3}
\nonumber \\
&-&\frac{2 \la \dot{\widehat{n}}(t')|\ddot{m}(t')\ra
+\la \widehat{n}(t')|\dddot{m}(t')\ra}{[i\omega_{nm}(t')]^3}
\nonumber \\
&+&  \frac{[\la \dot{\widehat{n}}(t')|\dot{m}(t')\ra + \la \widehat{n}(t')|\ddot{m}(t')\ra] 3i \dot{\omega}_{nm}(t')}
{[i\omega_{nm}(t')]^4}
\nonumber \\
&+& \frac{\la \widehat{n}(t')|\dot{m}(t')\ra i  \ddot{\omega}_{nm}(t')}{[i\omega_{nm}(t')]^4}
\nonumber \\
&+& \frac{\la \widehat{n}(t')|\dot{m}(t')\ra 3 \dot{\omega}_{nm}^2(t')}{[i\omega_{nm}(t')]^5}  \Bigg\}  e^{iW_{nm}(t')}  \Bigg|_0^t
\nonumber \\
&+& \int_0^t v_{2,n} du_{2,n}.
\nonumber
\eeqa
Further integration by parts of the reminders generates a series with increasing powers of $\omega_{nm}(t)$ 
in the denominators. Similarly, the change $s=t/t_f$ and writing derivatives and integrals with respect to $s$ 
provides a series in inverse powers of $t_f$.   

\section{Complex time analysis\label{apB}}
Figures \ref{fig_g,d} (a) and \ref{CPRbad} (a) for CPR demonstrate that the approximation
$g_+(t)\approx -(uv)_+(t)$ for the initially unoccupied state   
using integration by parts may be valid or it may fail. The approximation relies on  the contribution to the integral near the boundary time $t_f$, 
so it fails when other critical points become important, as in Fig. \ref{CPRbad} (a).  
Consider the integral in Eq. (\ref{aco}) rewritten as -$\int h(t') e^{\Phi(t')} dt'$, with $h(t')=\la \widehat{+}(t')|\partial_{t} {-}(t')\ra$ and
$\Phi(t')=iW_{+-}(t')$.   
%\beq
%\label{aco}
%g_n(t) =  -\int_0^t  \la \widehat{n}(t')|\dot{m}(t')\ra e^{iW_{nm}(t')} dt',
%\eeq
% 
Figures \ref{CPR8} (a,b) show the degeneracy points, $E_+(t)=E_-(t)$, for both cases in the complex time plane. They are   
branch cuts of the exponent $\Phi$, see Figs. \ref{CPR8} (c,d),  and in addition poles of the 
function $h$. The original integral goes along the real axis. The two cases studied correspond to two very different 
configurations of the function $\Phi$ in the complex time plane, as shown in Figs. \ref{CPR8} (e-h).
When the approximation works, see Fig. 
\ref{fig_g,d} and the left column in Fig. \ref{CPR8}, Re($\Phi$) decreases towards the upper half-plane so that a steepest descent path from $t_f$, almost perpendicular to the
real axis, see Fig. \ref{CPR8} (e),
provides the dominant contribution to the integral.  A path towards $t=0$ can be drawn through the valley without giving any 
significant contribution to the integral, see Fig. \ref{CPR8} (g). 
When the approximation fails, see Fig. \ref{CPRbad}  and the right column in Fig. \ref{CPR8}, 
a steepest descent path goes from $t_f$ to $t=0$ along the real axis, see Fig. \ref{CPR8} (f). 
Upper and lower degenerate points are 
now at very similar heights, see Fig. \ref{CPR8} (h). Re($\Phi$) decreases monotonously along the real line towards $t=0$, but now the close  singularities of the function 
$h$ imply a strong disturbance and contribution around $t_f/2$.   
%%%%%%%%%%%%%%%%%%%%%%%%%%%%%%%%%%
\begin{figure}[h]
\begin{center}
\includegraphics[height=2.5cm,angle=0]{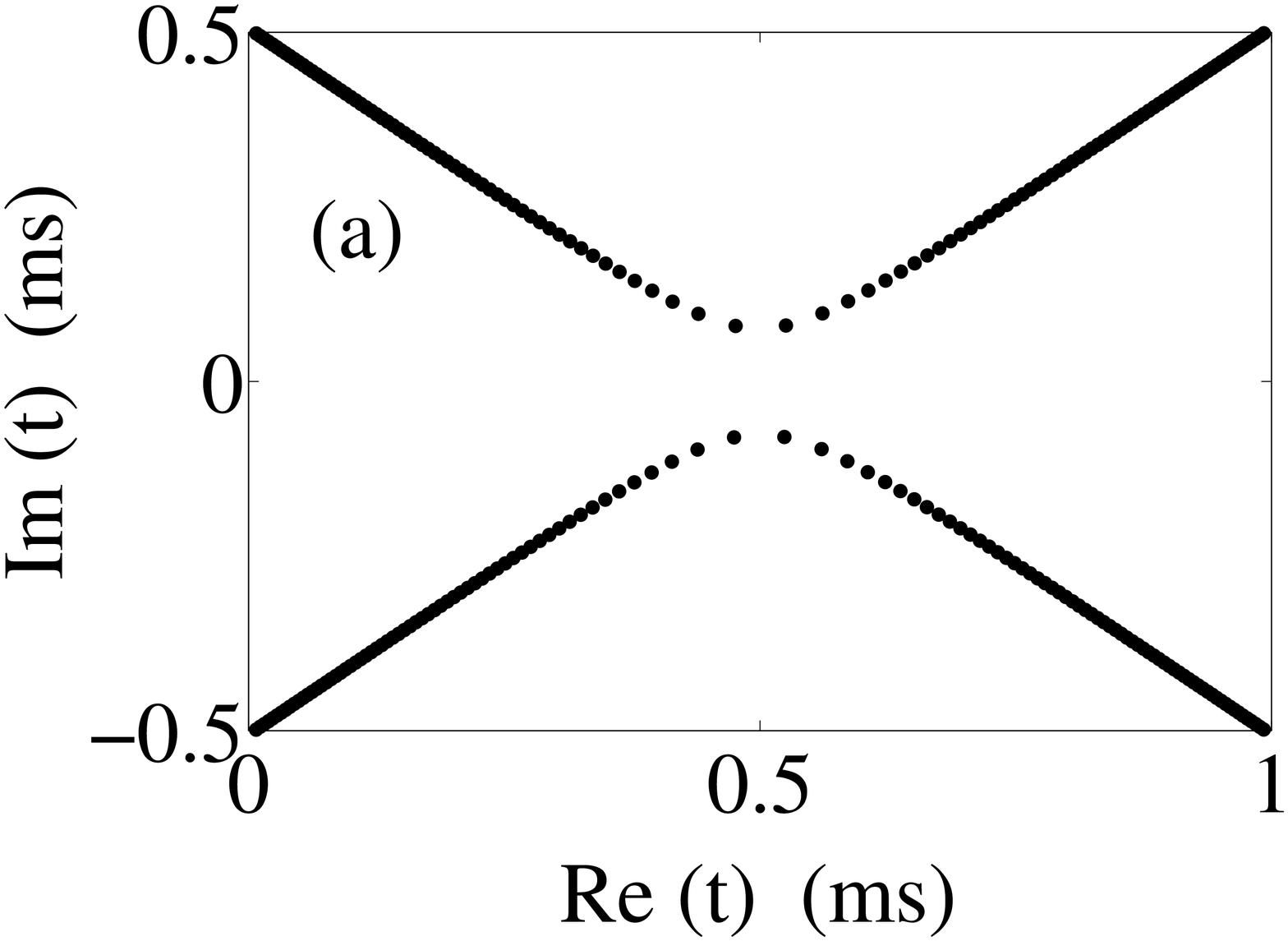}
\includegraphics[height=2.5cm,angle=0]{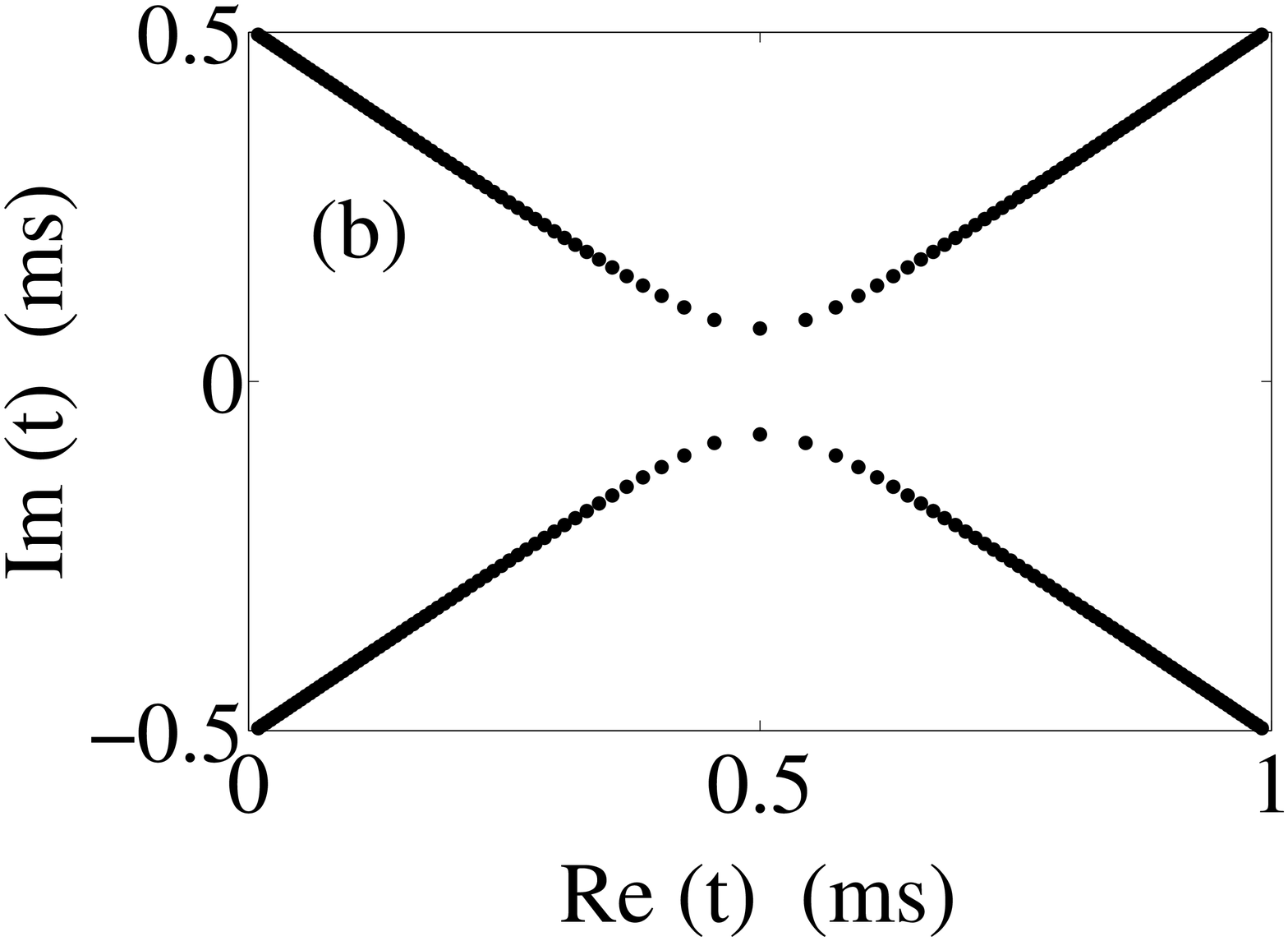}
\includegraphics[height=2.5cm,angle=0]{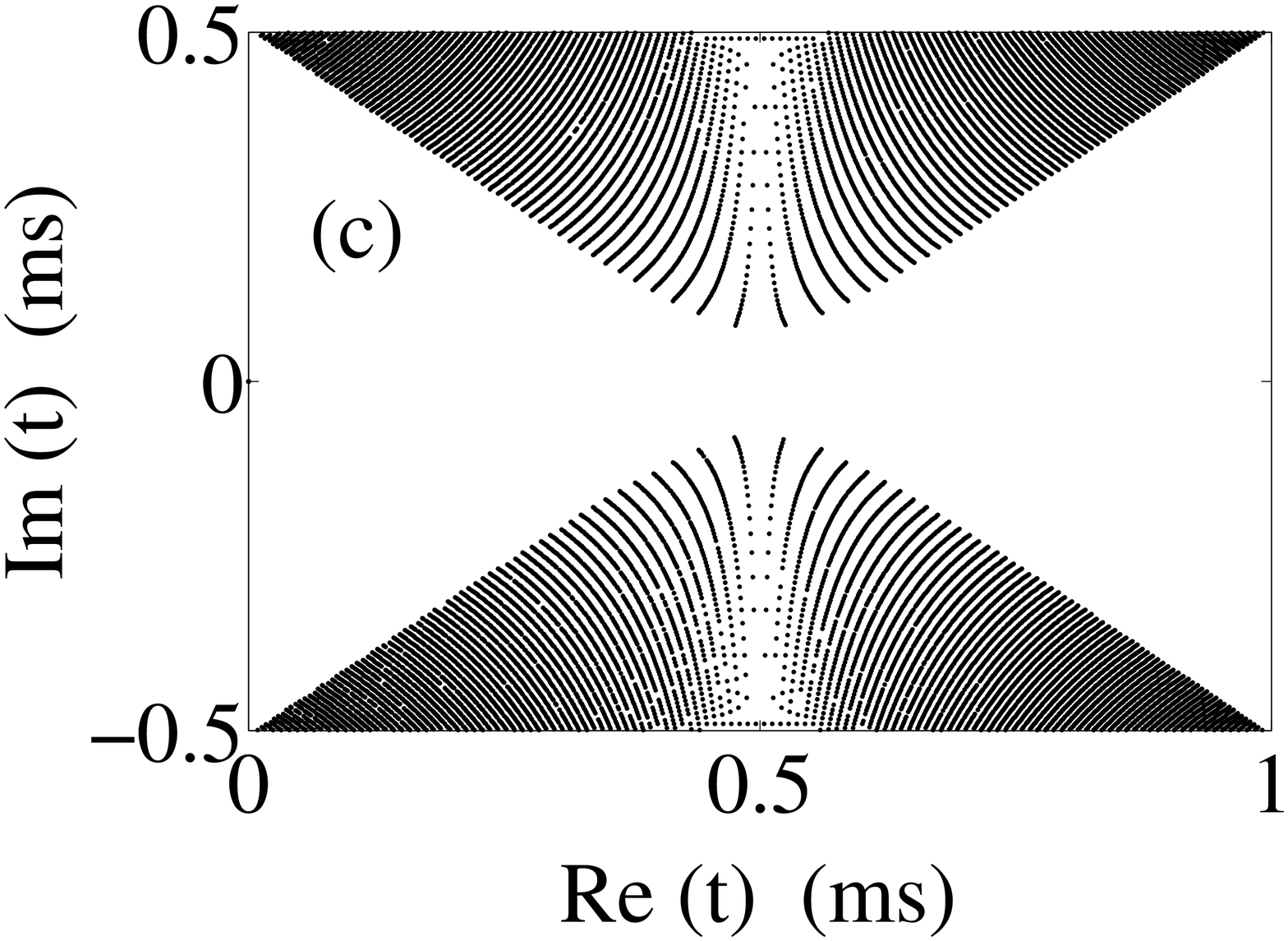}
\includegraphics[height=2.5cm,angle=0]{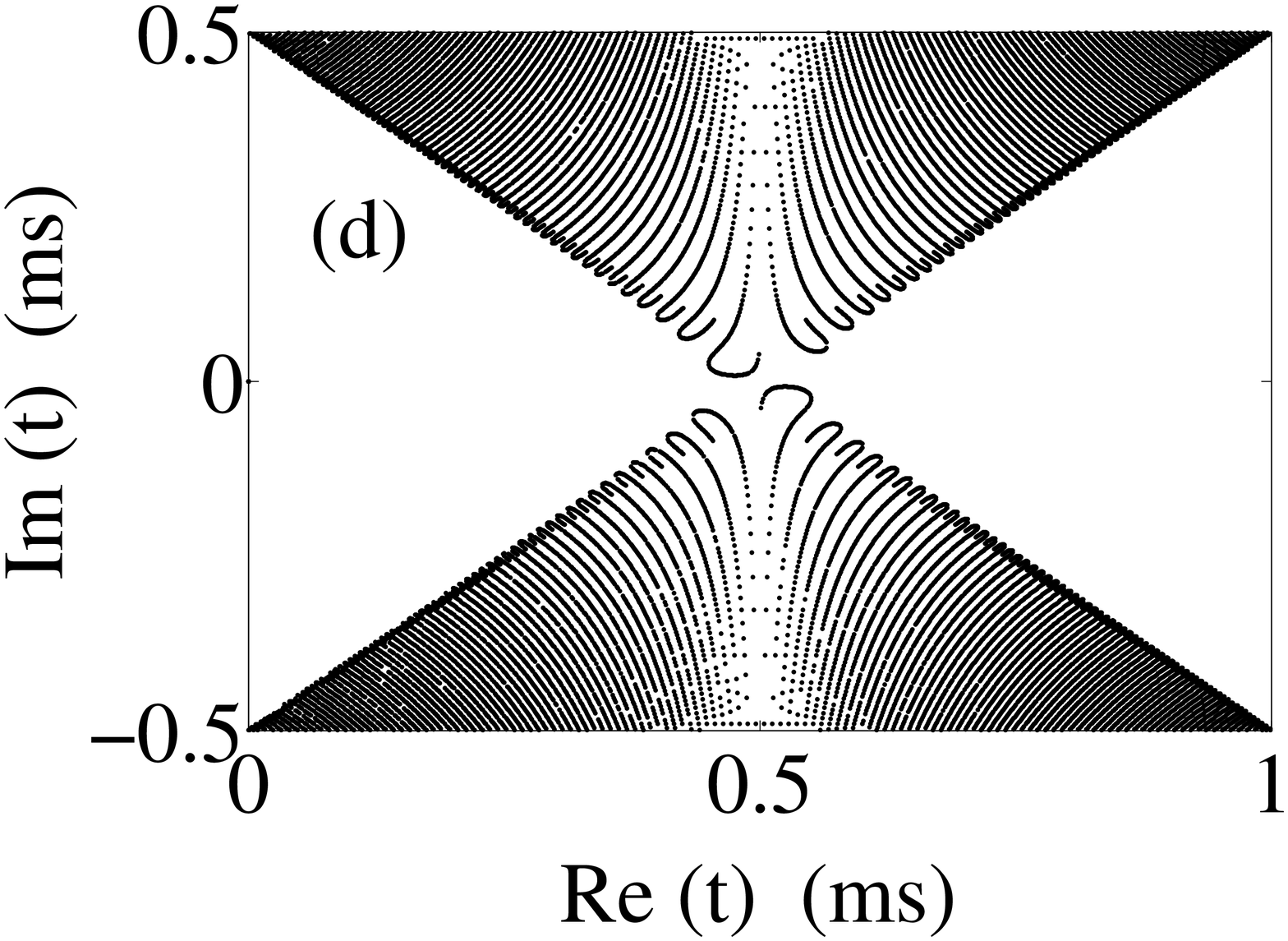}
\includegraphics[height=2.6cm,angle=0]{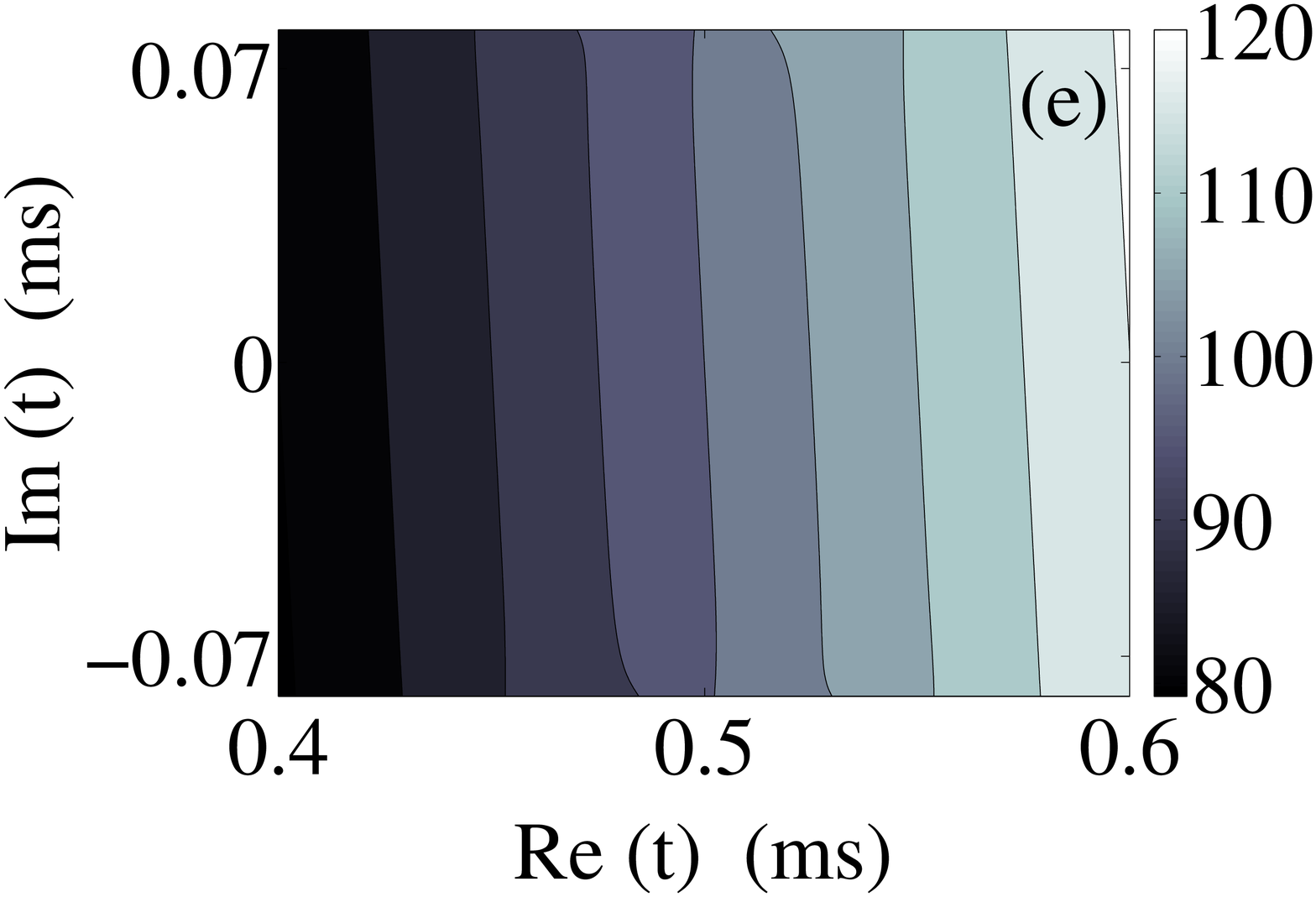}
\includegraphics[height=2.7cm,angle=0]{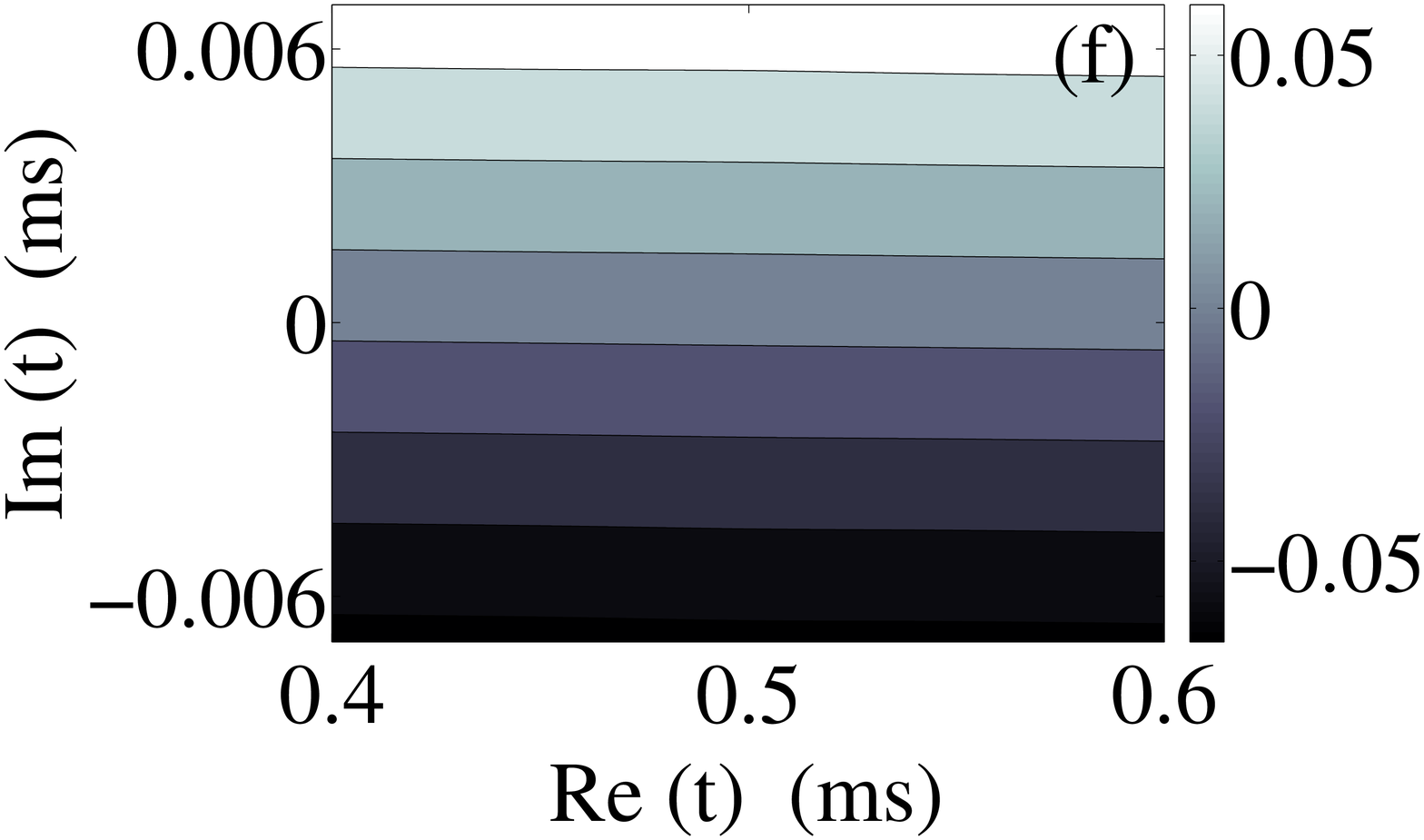}
\includegraphics[height=2.6cm,angle=0]{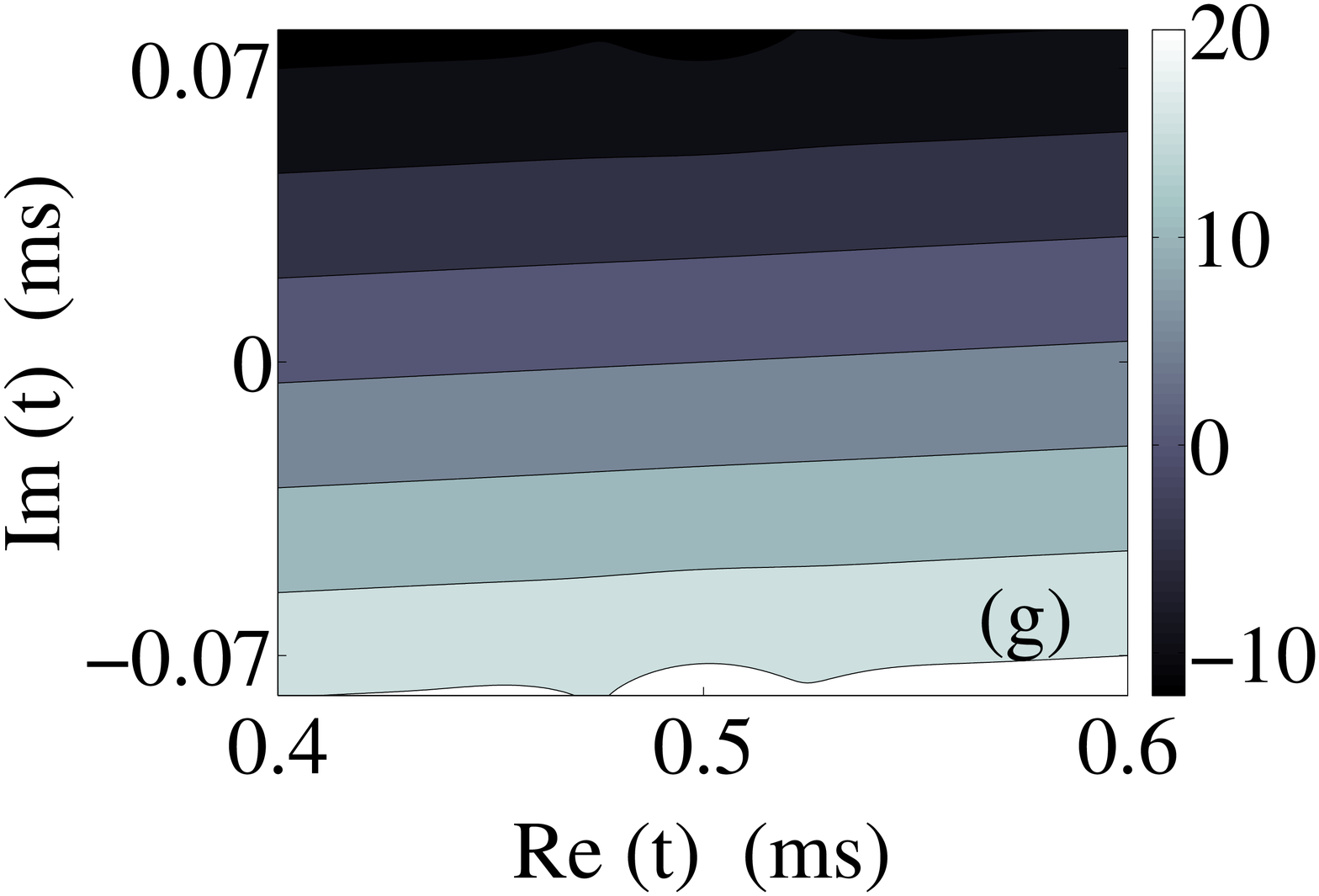}
\includegraphics[height=2.6cm,angle=0]{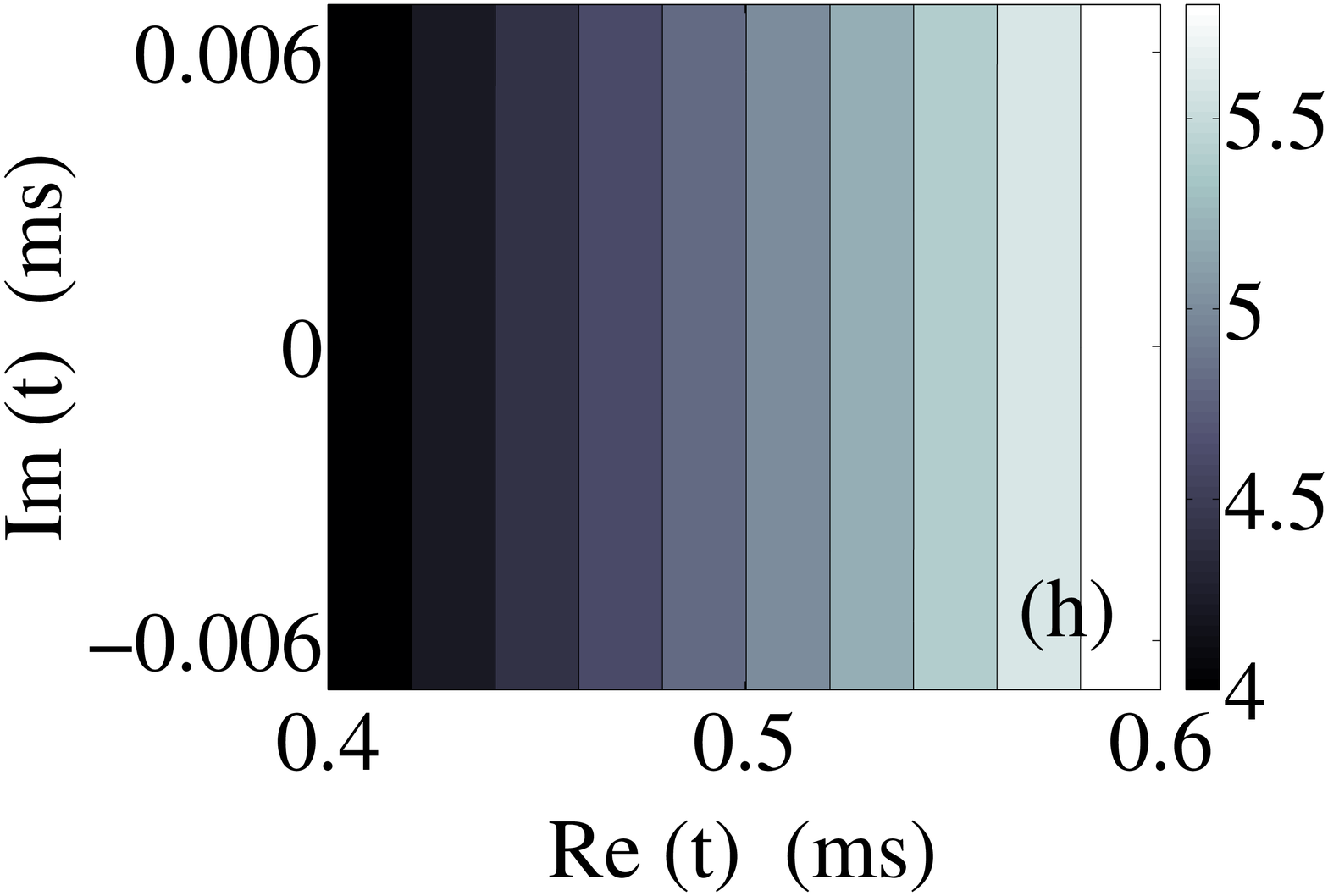}
\end{center}
\caption{\label{CPR8}
(Color online)
Complex time analysis of the integral for $g_+(t)$, see Eq. (\ref{aco}), for the CPR processes of Fig. \ref{fig_g,d} (left panels) 
and Fig. \ref{CPRbad} (right panels). (a,b) degeneracy points $(E_+=E_-)$; (c,d) branch cuts of the exponent
$\Phi$ following the criterion in  Fig. 3 (a);
(e,f) Im($\Phi$);  (g,h) Re($\Phi$). 
The contour maps  show a rectangle around $t_f/2$ not including the singularities. 
}
\end{figure}


\begin{thebibliography}{10}
%
%
\bibitem{BF} M. Born and V. Fock, Z. Phys. \textbf{51}, 165 (1928).
\bibitem{berry1984} M. V. Berry, Proc. R. Soc. Lond. A \textbf{392}, 45 (1984).
\bibitem{ad1} A. Ruschhaupt and J. G. Muga, Phys. Rev. A \textbf{70}, 061604(R) (2004). 
\bibitem{ad2}A. Ruschhaupt and J. G. Muga, Phys. Rev. A \textbf{73}, 013608 (2006). 
\bibitem{qc} E. Farhi, J. Goldstone, S. Gutmann, and M. Sipser, quant-ph/0001106. 
\bibitem{Schiff} L. I. Schiff, \textit{Quantum Mechanics}, McGraw-Hill, New York, 1981.
\bibitem{jolicard} A. Leclerc, D. Viennot, and G. Jolicard, J. Phys. A: Math. Theor. \textbf{45}, 415201 (2012).
\bibitem{BU} M. V. Berry and R. Uzdin, J. Phys. A: Math. Theor. \textbf{44}, 435303 (2011). 
\bibitem{erratum} S. Ib\'{a}\~{n}ez, S. Mart\'{i}nez-Garaot, Xi Chen, E. Torrontegui, and J. G. Muga, Phys. Rev. A \textbf{86}, 019901(E) (2012).
\bibitem{nenciu} G. Nenciu and G. Rasche, J. Phys. A: Math. Gen. \textbf{25}, 5741 (1992).

\bibitem{ChangPu} C. P. Sun, Phys. Scr. \textbf{48}, 393 (1993).
\bibitem{guerin} G. Dridi, S. Gu\'{e}rin, H. R. Jauslin, D. Viennot, and G. Jolicard, Phys. Rev. A \textbf{82}, 022109 (2010).
\bibitem{FM} A. Fleischer and N. Moiseyev, Phys. Rev. A \textbf{72}, 032103 (2005). 
\bibitem{sar_sof} S. Ib\'{a}\~{n}ez, S. Mart\'{i}nez-Garaot, Xi Chen, E. Torrontegui, and J. G. Muga, Phys. Rev. A \textbf{84}, 023415 (2011).
\bibitem{Jolicard2012} D. Viennot, A. Leclerc, G. Jolicard, and J. P. Killingbeck, J. Phys. A: Math. Theor. \textbf{45}, 335301 (2012). 
\bibitem{GM}I. Gilary and N. Moiseyev, J. Phys. B: At. Mol. Opt. Phys. \textbf{45}, 051002 (2012).
\bibitem{Muga} J. G. Muga, J. P. Palao, B. Navarro, and I. L. Egusquiza, Phys. Rep. \textbf{395}, 357 (2004).
\bibitem{Comparat} D. Comparat, Phys. Rev. A \textbf{80}, 012106 (2009). 
\bibitem{qac} C. Guo, Q.-H. Duan, W. Wu, and P.-X. Chen, Phys. Rev. A \textbf{88}, 012114 (2013).  
\bibitem{putterman} A. Kvitsinsky and S. Putterman, J. Math.  Phys. \textbf{32}, 1403 (1991).
\bibitem{wright} J. C. Garrison and E. M. Wright, Phys. Lett. A \textbf{128}, 177 (1988).
\bibitem{CPR} N. V. Vitanov, B. W. Shore, L. Yatsenko, K. B\"ohmer, T. Halfmann, T. Rickes, and K. Bergmann, 
Opt. Commun., \textbf{199}, 117 (2001). 
\bibitem{Lizuain} J. G. Muga, J. Echanobe, A. del Campo, and I. Lizuain, J. Phys. B \textbf{41}, 175501 (2008).
\bibitem{Pritchard} E. W. Streed, J. Mun, M. Boyd, G. K. Campbell, P. Medley, W. Ketterle, and D. E. Pritchard, Phys. Rev. Lett. \textbf{97}, 260402 (2006).
\bibitem{sara} S. Ib\'{a}\~{n}ez, A. Peralta Conde, David Gu\'{e}ry-Odelin, and J. G. Muga, Phys. Rev. A \textbf{84}, 013428 (2011).
\bibitem{DG} G. Dridi and S. Gu\'erin, J. Phys. A: Math. Theor. \textbf{45}, 185303 (2012). 
\bibitem{Guerin2002} S. Gu\'erin, S. Thomas, and H. R. Jauslin, Phys. Rev. A {\bf 65}, 023409 (2002). 
\bibitem{Berry89} M. V. Berry, Proc. R. Soc. Lond. A {\bf 422}, 7 (1989). 
\bibitem{MS} K.-P. Marzlin and B. C. Sanders, Phys. Rev. Lett.  \textbf{93}, 160408 (2004); 
S. Duki, H. Mathur, and O. Narayan, {\it ibid.} \textbf{97}, 128901 (2006);
J. Ma, Y. Zhang, E. Wang, and B. Wu, {\it ibid.} \textbf{97}, 128902 (2006); 
 K.-P. Marzlin and B. C. Sanders, {\it ibid.} \textbf{97}, 128903 (2006).
\bibitem{Duetal} J. Du, L. Hu, Y. Wang, J. Wu, M. Zhao, and D. Suter, Phys. Rev. Lett. \textbf{101}, 060403 (2008).  
\bibitem{PRLS} D. M. Tong, Phys. Rev. Lett. \textbf{104}, 120401 (2010);
M. S. Zhao and J. D. Wu, {\it ibid.}  \textbf{106}, 138901 (2011);
D. Comparat, {\it ibid.} \textbf{106}, 138902 (2011);
D. M. Tong,  {\it ibid.} \textbf{106}, 138903 (2011).
\bibitem{Ortigoso} J. Ortigoso, Phys. Rev. A \textbf{86}, 032121 (2012).
\bibitem{UMM} R. Uzdin, A. Mailybaev, and N. Moiseyev, J. Phys. A: Math. Theor. \textbf{44}, 435302 (2011). 
\bibitem{LGJ} X. Lacour, S. Gu\'erin, and H. R. Jauslin, Phys. Rev. A \textbf{78}, 033417 (2008). 
\bibitem{AT} A. C. Aguiar Pinto and M. T. Thomaz, J. Phys. A: Math. Gen. \textbf{36}, 7461 (2003). 
\bibitem{Joye} A. Joye, Commun. Math. Phys. \textbf{275}, 139 (2007). 
\bibitem{ICM} S. Ib\'{a}\~{n}ez, X. Chen, and J. G. Muga, Phys. Rev. A \textbf{87}, 043402 (2013).  
\bibitem{NB} A. I. Nesterov and G. P. Berman, Phys. Rev. A \textbf{86}, 052316 (2012). 
\bibitem{TDL} B. T. Torosov, G. Della Valle, and S. Longhi, Phys. Rev. A \textbf{87}, 052502 (2013).
 
%\bibitem{NMR} F. Bloch, Phys. Rev. \textbf{70}, 460 (1946).

 
\end{thebibliography}
\end{document}